%% file: constructive_ucode.tex
\documentclass[sigconf]{acmart}

\input{custom}

\begin{document}

\copyrightyear{2018}
\acmYear{2018}
\setcopyright{acmlicensed}
\acmConference[CCS '18]{2018 ACM SIGSAC Conference on Computer and Communications Security}{October 15--19, 2018}{Toronto, ON, Canada}
\acmBooktitle{2018 ACM SIGSAC Conference on Computer and Communications Security (CCS '18), October 15--19, 2018, Toronto, ON, Canada}
\acmPrice{15.00}
\acmDOI{10.1145/3243734.3243861}
\acmISBN{978-1-4503-5693-0/18/10}

\title{An Exploratory Analysis of Microcode \\as a Building Block for System Defenses}

\author{Benjamin Kollenda, Philipp Koppe, Marc Fyrbiak \\Christian Kison, Christof Paar, Thorsten Holz}
\affiliation{\institution{Ruhr-Universit\"at Bochum}}
\email{firstname.lastname@rub.de}

\input{sections/abstract}

\ccsdesc[500]{Security and privacy~Systems security}
\ccsdesc[300]{Security and privacy~Software and application security}

\keywords{security, microcode, defense}

\maketitle

\input{sections/introduction}

\input{sections/background}

\input{sections/reversing}

\input{sections/primitives}

\input{sections/discussion}

\input{sections/conclusion}

\urlstyle{tt}
\bibliographystyle{ACM-Reference-Format}
\bibliography{bibliography}

\input{sections/appendix}

\end{document}

%% file: custom.tex
\usepackage[english]{babel}
\usepackage{lipsum}
\usepackage{caption}
\usepackage{subcaption}
\usepackage{graphicx}
\usepackage{url}
\makeatletter
\g@addto@macro{\UrlBreaks}{\UrlOrds}
\makeatother
\PassOptionsToPackage{hyphens}{url}
\PassOptionsToPackage{hidelinks}{hyperref}

\usepackage{listings}
\usepackage{booktabs}
\usepackage{array}
\usepackage{multirow,tabularx}
\usepackage{makecell}
\usepackage{paralist}
\usepackage{balance}
\usepackage{xspace}
\usepackage{amsmath}
\usepackage{tikz}
\usetikzlibrary{shadows,arrows,shapes}
\usepackage{comment}

\usepackage{lscape}

\usepackage{moresize}

\usepackage{anyfontsize}

\usepackage{marginnote}

\usepackage{pifont}

\usepackage{colortbl}

\usepackage{verbatimbox}

\usepackage{xspace}

\usepackage{microtype}

\usepackage[nolist]{acronym}
\input{acronym}

\pdfmapfile{+zi4.map}
\usepackage{zi4}

\newcolumntype{L}[1]{>{\raggedright\let\newline\\\arraybackslash\hspace{0pt}}m{#1}}
\newcolumntype{C}[1]{>{\centering\let\newline\\\arraybackslash\hspace{0pt}}m{#1}}
\newcolumntype{R}[1]{>{\raggedleft\let\newline\\\arraybackslash\hspace{0pt}}m{#1}}

\lstdefinelanguage{JavaScriptHTML}{
    keywords={typeof, new, true, false, catch, function, return, null, catch, switch, var, if, in, while, do, else, case, break, function},
    keywordstyle=\color{BrickRed}\bfseries,
    ndkeywords={class, export, boolean, throw, implements, import, this},
    ndkeywordstyle=\color{Gray}\bfseries,
    identifierstyle=\color{black},
    sensitive=false,
    comment=[l]{//},
    morecomment=[s]{/*}{*/},
    morecomment=[s]{<!--}{-->},
    commentstyle=\color{Green}\ttfamily,
    stringstyle=\color{red}\ttfamily,
    morestring=[b]',
    morestring=[b]"
}

\lstdefinestyle{JS}{language=JavaScriptHTML,
    extendedchars=true,
    basicstyle=\scriptsize\ttfamily,
    showstringspaces=false,
    showspaces=false,
    numberstyle=\tiny\color{black},
    numbers=left,
    numbersep=8pt,
    tabsize=2,
    breaklines=true,
    showtabs=false,
    frame=single,
    frameround=ffff,
    captionpos=b
}

\newcommand{\etal}{et\,al.\xspace}
\newcommand{\eg}{e.\,g.\xspace}

%% file: acronym.tex
\begin{acronym}

\acro{3DES}{Triple-DES}

\acro{AES}{Advanced Encryption Standard}
\acro{ALU}{Arithmetic Logic Unit}
\acro{AOT}{Ahead-of-Time}
\acro{API}{Application Programming Interface}
\acro{ARX}{Addition Rotation XOR}
\acro{ASIC}{Application Specific Integrated Circuit}
\acro{ASIP}{Application Specific Instruction-Set Processor}
\acro{AS}{Active Serial}
\acro{ASAN}{Address Sanitizer}
\acro{HWASAN}{Address Sanitizer in Hardware}

\acro{BNF}{Backus-Naur Form}
\acro{BRAM}{Block-Ram}
\acro{BIOS}{Basic Input/Output System}

\acro{CBC}{Cipher Block Chaining}
\acro{CET}{Control-Flow Enforcement Technology}
\acro{CFB}{Cipher Feedback Mode}
\acro{CFG}{Control Flow Graph}
\acro{CFI}{Control Flow Integrity}
\acro{CISC}{Complex Instruction Set Computer}
\acro{CLB}{Configurable Logic Block}
\acro{CMP}{Chemical Mechanical Polishing}
\acro{COFF}{Common Object File Format}
\acro{COTS}{Commercial Off-The-Shelf}
\acro{CPA}{Correlation Power Analysis}
\acro{CPI}{Code-Pointer Integrity}
\acro{CPU}{Central Processing Unit}
\acro{CRC}{Cyclic Redundancy Check}
\acro{CTR}{Counter}

\acro{DC}{Direct Current}
\acro{DCFG}{Dynamic Control Flow Graph}
\acro{DES}{Data Encryption Standard}
\acro{DFA}{Differential Frequency Analysis}
\acro{DFT}{Discrete Fourier Transform}
\acro{DLL}{Dynamic Link Library}
\acro{DMA}{Direct Memory Access}
\acro{DNF}{Disjunctive Normal Form}
\acro{DPA}{Differential Power Analysis}
\acro{DRAM}{Dynamic Random Access Memory}
\acro{DSO}{Digital Storage Oscilloscope}
\acro{DSP}{Digital Signal Processing}
\acro{DUT}{Device Under Test}
\acro{DRAM}{Dynamic random-access memory}

\acro{ECB}{Electronic Code Book}
\acro{ECC}{Elliptic Curve Cryptography}
\acro{ECDH}{Elliptic Curve Diffie-Hellman}
\acro{EEPROM}{Electrically Erasable Programmable Read-only Memory}
\acro{EMA}{Electromagnetic Emanation}
\acro{EM}{electro-magnetic}

\acro{FFT}{Fast Fourier Transformation}
\acro{FF}{Flip Flop}
\acro{FI}{Fault Injection}
\acro{FIR}{Finite Impulse Response}
\acro{FPGA}{Field Programmable Gate Array}
\acro{FSM}{Finite State Machine}

\acro{GUI}{Graphical User Interface}

\acro{HDL}{Hardware Description Language}
\acro{HD}{Hamming Distance}
\acro{HF}{High Frequency}
\acro{HSM}{Hardware Security Module}
\acro{HW}{Hamming Weight}
\acro{HMAC}{Hash-based message authentication code}
\acro{HAFIX}{Hardware-Assisted Flow Integrity eXtension}

\acro{IC}{Integrated Circuit}
\acro{IDU}{Instruction Decode Unit}
\acro{IO}[I/O]{Input/Output}
\acro{IOB}{Input Output Block}
\acro{IOT}[IoT]{Internet of Things}
\acro{IP}{Intellectual Property}
\acro{IR}{Instruction Register}
\acro{ISA}{Instruction Set Architecture}
\acro{ISR}[ISR]{Instruction Set Randomization}
\acro{IV}{Initialization Vector}

\acro{JIT}{Just-in-Time}
\acro{JS}{JavaScript}
\acro{JTAG}{Joint Test Action Group}

\acro{KAT}{Known Answer Test}

\acro{LFSR}{Linear Feedback Shift Register}
\acro{LSB}{Least Significant Bit}
\acro{LUT}{Look-up table}

\acro{MAC}{Message Authentication Code}
\acro{MILS}{Multiple Independent Levels of Security}
\acro{MIPS}{Microprocessor without Interlocked Pipeline Stages}
\acro{MMIO}{Memory Mapped \acl{IO}}
\acro{MSB}{Most Significant Bit}
\acro{MSR}[MSR]{Model-specific register}

\acro{NASA}{National Aeronautics and Space Administration}
\acro{NSA}{National Security Agency}
\acro{NVM}{Non-Volatile Memory}

\acro{OFB}{Output Feedback Mode}
\acro{OISC}{One Instruction Set Computer}
\acro{ORAM}{Oblivious Random Access Memory}
\acro{OS}{Operating System}

\acro{PAR}{Place-and-Route}
\acro{PCB}{Printed Circuit Board}
\acro{PC}{Program Counter}
\acro{PLA}{Programmable Logic Array}
\acro{PS}{Passive Serial}
\acro{PUF}{Physical Unclonable Function}

\acro{RAM}{Random Access Memory}
\acro{SEM}{Scanning Electron Microscope}
\acro{RISC}{Reduced Instruction Set Computer}
\acro{RNG}{Random Number Generator}
\acro{ROM}{Read-Only Memory}
\acro{ROI}{Region Of Interest}
\acro{ROP}{Return-oriented Programming}
\acro{RTL}{Register Transfer Language}

\acro{SCA}{Side-Channel Analysis}
\acro{SHA}{Secure Hash Algorithm}
\acro{SNR}{Signal-to-Noise Ratio}
\acro{SPA}{Simple Power Analysis}
\acro{SPI}{Serial Peripheral Interface Bus}
\acro{SRAM}{Static Random Access Memory}
\acro{SGX}{Software Guard Extensions}
\acro{SOC}{System on a Chip}

\acro{TEA}[TEA]{Tiny Encryption Algorithm}
\acro{TSC}[TSC]{Time Stamp Counter}

\acro{UART}{Universal Asynchronous Receiver Transmitter}
\acro{UHF}{Ultra-High Frequency}
\acro{USB}{Universal Serial Bus}

\acro{UEFI}{Unified Extensible Firmware Interface}

\acro{VHDL}{Very High Speed Integrated Circuit Hardware Description Language}

\acro{WISC}{Writeable Instruction Set Computer}

\acro{XTS}{XEX-based Tweaked-codebook with ciphertext Stealing}
\acro{XoM}[XoM]{Execute-Only Memory}

\end{acronym}

%% file: sections/abstract.tex
\begin{abstract}

   Microcode is an abstraction layer used by modern x86 processors that interprets user-visible CISC instructions to hardware-internal RISC instructions. The capability to update x86 microcode enables a vendor to modify CPU behavior in-field, and thus patch erroneous microarchitectural processes or even implement new features. Most prominently, the recent \textsc{Spectre} and \textsc{Meltdown} vulnerabilities were mitigated by Intel via microcode updates. Unfortunately, microcode is proprietary and closed source, and there is little publicly available information on its inner workings. 

In this paper, we present new reverse engineering results that extend and complement the public knowledge of proprietary microcode. Based on these novel insights, we show how modern system defenses and tools can be realized in microcode on a commercial, off-the-shelf AMD x86 CPU. We demonstrate how well-established system security defenses such as timing attack mitigations, hardware-assisted address sanitization, and instruction set randomization can be realized in microcode. We also present a proof-of-concept implementation of a microcode-assisted instrumentation framework. Finally, we show how a secure microcode update mechanism and enclave functionality can be implemented in microcode to realize a small trusted execution environment. All microcode programs and the whole infrastructure needed to reproduce and extend our results are publicly available.

\end{abstract}

%% file: sections/introduction.tex
\section{Introduction}

New vulnerabilities, design flaws, and attack techniques with devastating consequences for the security and safety of computer systems are announced on a regular basis~\cite{cvestats}.
The underlying faults range from critical memory safety violations~\cite{cve:stats:memory} or input validation~\cite{cve:stats:validation} in software to race conditions or side-channel attacks in the underlying hardware~\cite{intel:errata,amd:errata,Kocher2018spectre,Lipp2018meltdown,projectzeromeltdownspectre,Hund,Doychev:2013:CTS}. To cope with erroneous behavior and to reduce the attack surface, various defenses have been developed and integrated in software and hardware over the last decades~\cite{van2017dynamics,Szekeres:2013:EWM}.

Generally speaking, defenses implemented in software can be categorized in either compiler-assisted defenses~\cite{ASAN,aslr,dep,onarlioglu2010g,ASLR-Guard,readactor:sp15,backes2014oxymoron} or binary defenses~\cite{wartell2012binary,pappas2012smashing,Gawlik,abadi2005control,Isomeron:ndss15}. Note that operating system changes~\cite{XnR:2014,readactor:sp15,aslr,dep} represent an orthogonal approach to serve both compiler-assisted and binary defenses. While compiler-assisted defenses require access to the source code and re-com\-pi\-la\-tion of the software, binary defenses based on \textit{static binary rewriting}~\cite{wang2015reassembleable,laurenzano2010pebil,romer1997instrumentation} or \textit{dynamic instrumentation}~\cite{dyninst:2011,luk2005pin,nethercote2007valgrind,dynamorio} can also be leveraged for legacy and \ac{COTS} programs. However, these binary defense strategies have two fundamental drawbacks: on the one hand, binary rewriting relies on the ability to accurately discover and disassemble all executable code in a given binary executable~\cite{andriesse2016depth}. Any misclassified code or data yields incomplete soundness and thus cannot provide specific security guarantees, causes program termination, or incorrect computations. On the other hand, dynamic instrumentation executes unmodified binaries and inserts instrumentation logic with methods such as \textit{emulation} or \textit{hooking} during runtime. While this approach does not require the availability of a perfect disassembly, it typically causes significant performance overheads and thus can be prohibitively expensive in practice.

Over the past decades, various defense mechanisms have been implemented in hardware to increase both security and performance. For example, dedicated security features to mitigate exploitation of memory-corruption vulnerabilities include Data Execution Prevention~\cite{dep}, \ac{XoM}~\cite{XnR:2014,readactor:sp15,intelsdm}, \ac{CFI}~\cite{abadi2005control,intel2016cet} and Shadow Stacks~\cite{intel2016cet,dang2015performance}. Moreover, sophisticated trusted computing security features were integrated in \acp{CPU}~\cite{anati2013innovative,iacr:2016:86}. 

But not only novel defense mechanisms have been integrated in hardware: Similarly to any complex software system, erratic behavior exist in virtually any commercially-available \ac{CPU}~\cite{intel:errata,amd:errata}. To this end, x86 \ac{CPU} vendors integrated in-field update features (e.g., to turn off defective parts or patch erroneous behavior). More precisely, the microcode unit, which translates between user-vis\-i\-ble \ac{CISC} \ac{ISA} and hardware-internal \ac{RISC} \ac{ISA}, can be updated by means of so-called \textit{microcode updates}~\cite{patent:2002:patch_device,koppe2017reverse}.
Since microcode is proprietary and closed source, and more and more complex security features are integrated into hardware with the help of microcode (e.g., Intel SGX~\cite{iacr:2016:86}), there is only a limited understanding of its inner workings and thus we need to trust the \ac{CPU} vendors that the security mechanisms are implemented correctly. In particular, the \ac{CPU}'s trustworthiness is challenged since even recently published microcode updates have been shown to cause incorrect behavior~\cite{intelspectreretract} and several attacks on hardware security features have been demonstrated recently~\cite{Kocher2018spectre,Lipp2018meltdown,projectzeromeltdownspectre,lee2017inferring,206170}. Moreover, since older \ac{CPU} generations are not updated to defend against sophisticated attacks such as \textsc{Spectre} or \textsc{Meltdown}~\cite{intelspectrepatch}, these \acp{CPU} are unprotected against the aforementioned attacks which find more and more adoption into real-world attacks~\cite{fortinetmdspecmw}.

\par{\bf Goals and Contributions.}
In this work, we focus on constructive applications of x86 processor microcode for the modern system security landscape. 
Our goal is to shed light on how currently employed defenses may be realized using microcode and thus tack\-le shortcomings of the opaque nature of x86 \acp{CPU}.
Building upon our recent work on microcode~\cite{koppe2017reverse}, we first present novel reverse engineering strategies which ultimately provide fine-grained understanding of x86 microcode for a \ac{COTS} AMD K8 \ac{CPU}. 
On this basis, we demonstrate multiple constructive applications implemented in microcode which considerably reduce the attack surface and simultaneously reduce performance overheads of software-only solutions. 
Finally, we discuss benefits and challenges for customizable microcode for future systems and applications. 

\smallskip \noindent
In summary, our main contributions are: 
\begin{itemize}

\item {\bf Uncovering New x86 Microcode Details.}
We present new reverse engineering results that extend and complement the publicly available knowledge of AMD K8 \ac{CPU} microcode technology, specifically its microcode \ac{ROM}. 
To this end, we develop a novel reverse engineering strategy that combines chip-level reverse engineering and image processing with a custom microcode emulator in order to recover and validate microcode semantics in a semi-automatic fashion. 
In particular, this reverse engineering step enables us to better understand the hitherto opaque microcode by analysis of its \ac{ROM} and microcode updates. %

\item {\bf Perspectives of Customizable Microcode.}
   We analyze the capabilities of microcode and its updates to identify building blocks that can be used to strengthen, extend, or supplement system security defenses. This includes microcode-based methods to enable or disable CPU features at runtime, a method to intercept low-level CPU processes, an isolated execution environment within the microcode engine, and the possibility to extend and modify the x86 \ac{ISA}.
With regards to the trustworthiness of systems, we discuss a method to detect the presence of microcode backdoors and the challenges associated with such a detection.

\item {\bf Implementation of Microcode-Assisted Defenses.}
We show how modern system defenses and tools can be implemented with microcode on a \ac{COTS} AMD x86 \ac{CPU} using the identified primitives. 
To this end, we implemented several case studies to demonstrate that timing attack mitigation, hardware-assisted address sanitization, and instruction set randomization can be realized in microcode. 
In addition, we realize a microcode-assisted hooking framework that allows fast filtering directly in microcode.
Finally, we show how a secure microcode update mechanism and enclave functionality can be implemented in microcode.
The framework used for the deconstruction
      and manipulation of microcode, including the assembler and disassembler, as well as our created microcode programs and the microcode emulator are publicly available at \url{https://github.com/RUB-SysSec/Microcode}~\cite{microcode:amd_microprograms}.
\end{itemize}

%% file: sections/background.tex
\section{Background and Related Work}

In the following, we first present the technical background information needed to understand the microcode details presented in this paper.
Note that the background for the individual defenses is covered in their respective subsections in Section~\ref{cucode:section:case_study}.
In addition, we review prior work that demonstrated the capabilities of microcode and discuss how our contributions presented in this paper relate to existing work.

\subsection{Microcode Background}

The \ac{ISA} of a processor defines the available instructions and serves as an interface
between software and hardware~\cite{book:computer_architecture:stallings}.
We refer to the actual hardware implementation of an \ac{ISA} as \textit{microarchitecture}.
The \ac{IDU} generates control words based on the currently decoded instruction and is a crucial
component of the microarchitecture especially for \ac{CISC} processors with complex instructions.
The \ac{IDU} of modern x86 processors is implemented as a hybrid of a hardwired decode unit,
which consists of sequential logic,
and a microcoded decode unit,
which replays precomputed control words named \textit{microinstructions}.
They are stored in a dedicated,
on-chip microcode \ac{ROM}.
The microcode is organized in so-called \textit{triads} containing three microinstructions and a \textit{sequence word},
which denotes the next triad to execute.
In the microcode address space, triads can only be addressed as a whole, i.e.,
individual bytes are not accessible.
There are multiple categories of microinstructions like arithmetic, logic,
memory load/store, and special microinstructions.

The microcode of modern x86 processors can be updated at runtime in order to fix errata and add new features
without the need to resort to product recalls~\cite{patent:2002:patch_device,koppe2017reverse}.
These updates are usually applied early during boot by the BIOS/EFI or operating system.
The process is initiated by loading the microcode update file to
main memory and writing the virtual address to a \ac{MSR}.
The CPU then copies the microinstructions of the update to the dedicated on-chip microcode \ac{RAM}.
The update engine also sets the \textit{match registers} according to the values given in the update file.
The match registers contain microcode \ac{ROM} addresses and act as breakpoints.
They redirect control to the triads of the update stored inside the
on-chip \ac{RAM} once a breakpoint in microcode \ac{ROM} is hit.
Complex or rarely used x86 instructions are implemented with
microcode and have a predefined entry point in microcode \ac{ROM}.
Hence,
microcoded x86 instructions can be intercepted by placing a breakpoint at the corresponding entry point.
The triads in the microcode update define the new logic of the x86 instruction.

\subsection{Related Work}

\par{\bf Microcode and Microcode Updates.}
Previous work~\cite{tr:2014:chen,hawkesmicrocode,link:2004:opteron_exposed} already provided indicators that the microcode
update functionality of several \acp{CPU} families is not sufficiently protected and might allow for custom updates to be applied.
Koppe~\etal~\cite{koppe2017reverse} then reverse engineered both the update mechanism of AMD K8 and K10 \acp{CPU} as
well as the encoding of microcode to a point that allowed the creation of custom microcode updates.
These updates implemented simple microcode applications such as basic instrumentation and backdoors,
which were applicable to unmodified \ac{CPU}.
Other work highlighting the capabilities of microcode was presented by Triulzi~\cite{Arrigo:2016:Troopers,Arrigo:2015:Troopers},
but details of the implementation are not publicly available.

In this paper, we substantially extend on these insights and perform further in-depth reverse engineering and analysis of the microcode \ac{ROM}. By understanding the \ac{ROM} mapping, we are able to disassemble the microcode of arbitrary x86 instructions to enable the implementation of sophisticated microprograms, as demonstrated in later sections of this work.

\par{\bf Microcoded Shadow Stacks.}
Davi~\etal~\cite{davi2015hafix} introduced an approach called \ac{HAFIX} and showed that it is possible to implement a so-called \textit{shadow stack}~\cite{dang2015performance} using microcode (in cooperation with researchers from Intel).
However, \ac{HAFIX} relied both on a compile-time component to add additional instructions to the binary,
and is only available on development \acp{CPU},
not on standard consumer hardware.
Intel also announced the introduction of shadow stacks into end user \acp{CPU} with the addition of
\ac{CET}~\cite{intel2016cet}.
This technology tracks all calls and returns which allows checking whether the normal stack and the shadow stack point to
the same return address.
If a difference is encountered,
an exception is raised.
Additionally, the memory pages containing the shadow stacks are protected using special page table attributes.
Once \acp{CPU} with this technology will reach the market,
shadow stacks will be available in production code with (almost) no additional performance overhead.

In this paper, we present several designs and proof-of-concept implementations of microcode-assisted systems defenses beyond shadow stacks. In addition, our paper and the supplementary material~\cite{microcode:amd_microprograms} will enable other researchers to build similar microcode-based system defenses and explore this area further.

%% file: sections/reversing.tex
\section{Microcode Reverse Engineering}

A key contribution of our work presented in this paper is to further analyze the \ac{ROM}
readouts provided by Koppe~\etal~\cite{koppe2017reverse} to gather more details on the
implementation of both microcode itself and---more importantly---on the microcoded instructions.
While the authors were able to identify operations and triads in the readout,
they were unable to reconstruct how they map to logical addresses.
Therefore,
they could not locate and analyze the microcode that implements a \emph{specific} x86 instruction.
However,
these steps are crucial in the hooking of more advanced x86 instructions that
require knowledge of the underlying implementation in the microcode \ac{ROM}.
The analysis of existing microcode implementations was essential for
the case studies presented in Section~\ref{cucode:section:case_study}.

\begin{figure}[t]
	\centering
	\resizebox{\linewidth}{!}{
		\includegraphics[scale=0.01]{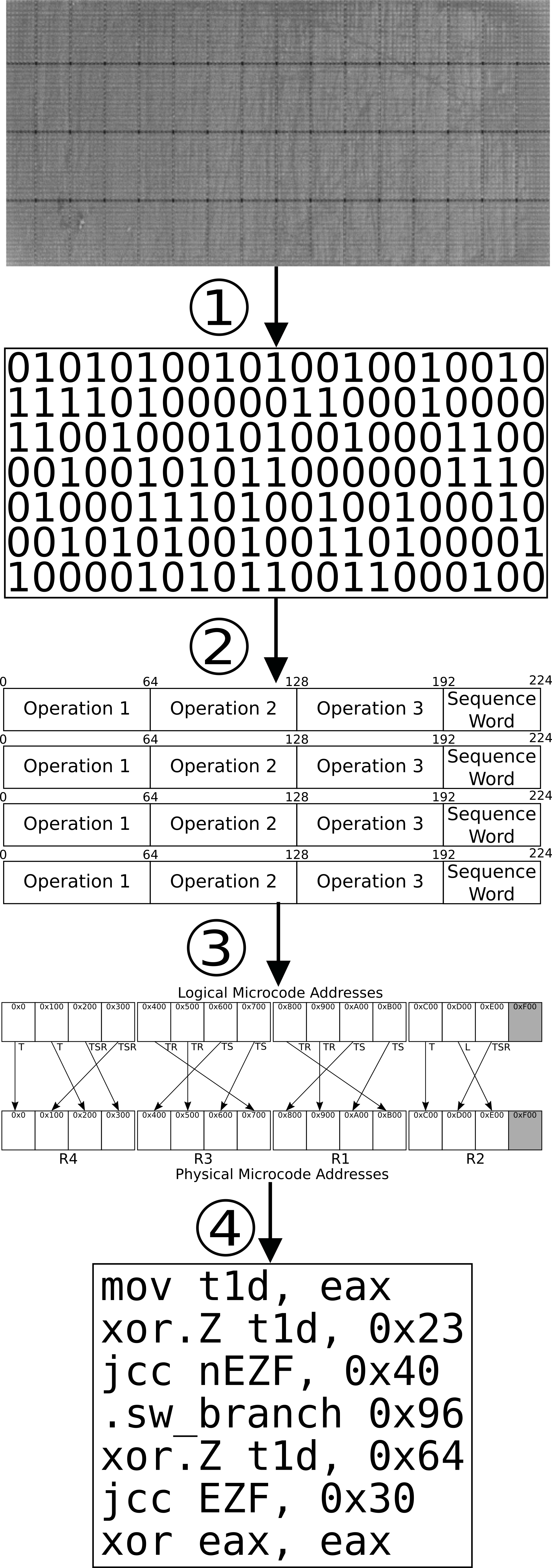}
	}
	\caption{High-level overview of the individual steps of the \ac{ROM} reverse engineering process.}
	\label{cucode:figure:rom:re_process}
\end{figure}

The key requirement for such an analysis is the ability to locate
the corresponding implementation in the microcode \ac{ROM}.
We therefore require a mapping of observable addresses to the physical location in the \ac{ROM} readout.
Going forward,
we define two different classes of addresses:
\begin{itemize}
   \item \emph{logical addresses} are used when the microcode software refers to a specific triad (\eg, in the match registers or jumps)
	\item \emph{physical addresses} are the addresses assigned to triads in the ROM readouts during analysis.
\end{itemize}
These addresses are not related to the virtual and physical addresses used when addressing the main memory---what is
commonly known as the \emph{virtual memory layout} of processes.
Also note that the address granularity for microcode is one triad,
the individual operations forming a triad are not addressable.

Thus, it is our goal to reverse engineer the algorithm used to map a given logical address to its corresponding physical
address.
The high level overview of this process is illustrated in Figure~\ref{cucode:figure:rom:re_process}.
We used the following steps to recover the ordered microcode \ac{ROM}:
\begin{itemize}
	\item \raisebox{.5pt}{\textcircled{\raisebox{-.9pt} {1}}}~Convert \ac{SEM} images
	of each region to bitstrings with the aid of image recognition software.
	\item \raisebox{.5pt}{\textcircled{\raisebox{-.9pt} {2}}}~Reorder
	and combine the resulting bitstrings into a list of unordered triads.
	\item \raisebox{.5pt}{\textcircled{\raisebox{-.9pt} {3}}}~Reconstruct the mapping between logical
	and physical microcode addresses as well as reorder the triads according to this mapping.
	\item \raisebox{.5pt}{\textcircled{\raisebox{-.9pt} {4}}}~Disassemble the resulting triad list into a continuous,
	ordered stream of instructions.
\end{itemize}
The first step,
the conversion of images to bitstrings,
was already performed by Koppe~\etal~\cite{koppe2017reverse} and we used this data as our starting point for our further analysis.
The authors also already combined parts of the readouts into triads.
We build upon this and recovered the remaining part of the triads,
which is depicted as step \raisebox{.5pt}{\textcircled{\raisebox{-.9pt} {2}}}~in the figure.
The details of this step are described in
Sections~\ref{cucode:section:re:layout}
and~\ref{cucode:section:re:ordering}.
Step \raisebox{.5pt}{\textcircled{\raisebox{-.9pt} {3}}},~the recovery of the mapping algorithm,
constituted the majority of our efforts.
We outline the approach we used in Section~\ref{cucode:section:re:approach}
and provide details of the solutions we developed in the following sections.
The mapping was reverse engineered for an AMD K8 processor.
However, our approach is also applicable to the K10 architecture based on the
similarities between the two architectures.
For the last step, we extended the disassembler used by Koppe~\etal~\cite{koppe2017reverse} to include details learned during our own analysis.

\begin{figure}[!t]
	\centering
	\resizebox{\linewidth}{!}{
		\includegraphics{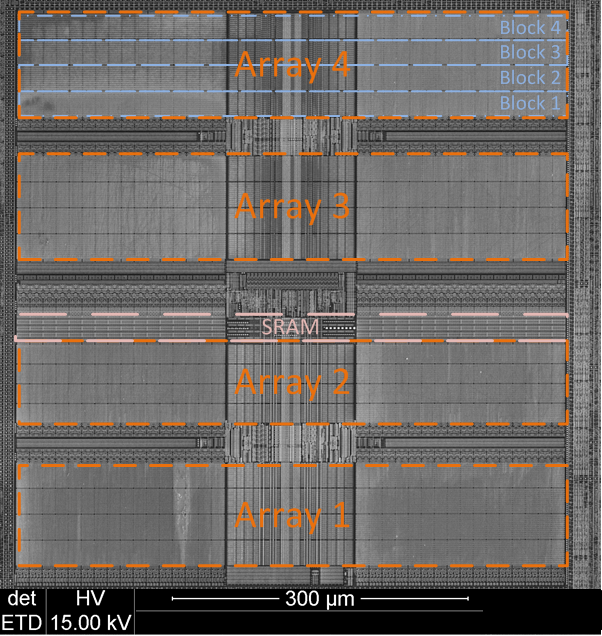}
	}
	\caption{\ac{SEM} image of region R1 showing arrays A1 to A4 and the SRAM holding the microcode update. The higher resolution raw image is available in Appendix~\ref{cucode:section:appendix:rom}.}
	\label{cucode:figure:rom:region}
\end{figure}

\subsection{Physical Layout}
\label{cucode:section:re:layout}
The physical storage is composed of three larger regions of \ac{ROM} (R1 to R3),
which were identified as the area containing the operations,
and a smaller region (R4) containing the sequence words.
Previous work~\cite{koppe2017reverse} already performed permutations such as inversion and interleaving of bit rows to receive whole operations in the correct bit order.
In addition, the algorithm for constructing triads out of three operations was known.
The triads are built by loading a single operation out of each of the three regions R1 to R3 and loading the corresponding
sequence word from region R4. Thereby, the operations belonging to one triad have the same offset relative to the start of their corresponding region.
The different subregions of a single \ac{ROM} region are illustrated in Figure~\ref{cucode:figure:rom:region}, more technical details are provided in Appendix~\ref{cucode:section:appendix:rom}.
We will use the same naming convention in the following.

The hardware layout suggested that the triads are organized in four arrays (A1 to A4), with A1,
A3 and A4 containing data for 1024 triads each and A2,
which is physically smaller than the other arrays,
for 768 triads.
This organization means that the first triad
will use bits extracted from R1:A1,
R2:A1 and R3:A1 as its operations and the sequence word is obtained from the bits located in R4:A1.
As the regions are no longer relevant after combining the triads,
they will be omitted in further notations.
Each of the arrays is subdivided into blocks B1 to B4, each containing 256 triads.
The exception to this is the array A2:
while the hardware layout suggests the presence of four blocks with a smaller number of triads each,
we mapped the contents to three blocks with 256 triads each.
This means array A2 contains only 768 triads in contrast to the 1024 triads contained in the other arrays.

We were also able to locate the microcode
patch \ac{RAM},
which is loaded with the microcode updates during runtime.
The \ac{RAM} needs to be placed physically close to the rest of the microcode engine to keep signal paths short,
however previously it was unknown where exactly it is located.
Using new images taken with a \ac{SEM}, we could classify the area between arrays A2 and A3 as \ac{SRAM}.
The area is marked in Figure~\ref{cucode:figure:rom:region}.
We determined the storage type based on detailed images of the region
and additional cross-section images.
Both showed visual structures specific to \ac{SRAM}.
This location also contains a visually different control logic,
which also indicates a different type of storage than the rest of the region.
A higher resolution image and additional details are available in Appendix~\ref{cucode:section:appendix:rom}.
It should be noted that the usage of two different classes of storage in this close proximity implies a highly optimized hardware layout.
The SRAM marked in the figure contains $32\times64$ bits,
which is the amount of data needed per region for 32 triads.
This corresponds to the maximum update size of 32 triads determined in our experiments.
Due to the additional complexity of implementing a fast readable and writable memory in hardware,
the \ac{SRAM} occupies roughly the same space as a \ac{ROM} block with 256 triads.

\subsection{Physical Ordering}
\label{cucode:section:re:ordering}
Another insight gained from the available readout was that not only the three operations forming a triad exhibited data
dependencies between each other (suggesting that the triads are indeed correctly combined),
but in some cases data flow was visible between triads immediately following each other.
This means the readout already partially placed related triads close to each other.
Based on this observation, we retained the triad order and by convention placed all triads after one another with increasing
addresses.
This yielded what we considered a continuous physical memory space with addresses starting at 0 and increasing with each
triad to 0xEFF.
This corresponded with the observation that the microcode patch RAM starts at the address 0xF00 for the K8 series of
processors.

Our physical memory space assumed an arbitrary ordering of A1 -- A3 -- A4 -- A2,
so A1 would contain addresses from 0x0 to 0x3FF,
A3 from 0x400 to 0x7FF,
A4 from 0x800 to 0xBFF and A2 from 0xC00 to 0xEFF.
We placed A2 last
because it contained less triads which we assumed to be missing at the end of the addressable space.
In each array, we ordered the blocks starting from the bottom of the image in Figure~\ref{cucode:figure:rom:region},
omitting the missing block B4 in array A2.
Physical address 0x0 is thus located in A1:B1 and 0xEFF in A2:B3.

\subsection{Mapping Recovery Approach}
\label{cucode:section:re:approach}
Our recovery approach is based on inferring the mapping based on address pairs.
We chose this approach because it was infeasible to recover the mapping via hardware analysis.
The addressing logic is complex and the connections span multiple layers, each of which would require delayering and subsequent imaging.
Each address pair maps a logical (microcode) address to a physical address.
Once the recovered function correctly produces the physical address for any given logical address in our test set,
we can assume that it will be correct for any further translations.
We thus needed a sufficiently large collection of address pairs.
Unfortunately, the microcode updates only provided two usable data points.

Therefore, we developed an approach that (i) executes all \ac{ROM} triads on the \ac{CPU} individually and extracts the observable semantics of a given logical address, (ii) emulates each triad we acquired from the physical \ac{ROM} readout in a custom microcode emulator to extract the semantics for a given physical address, and (iii) correlates the extracted semantics to find matching pairs of physical and logical addresses. Details of this process are described in Section~\ref{cucode:section:re:emulation}. This resulted in a total of $54$ address pairs.
The results were then reviewed in a manual analysis step to find the correct permutation of triads for a given block.
Once a permutation candidate for a block is found, it can be verified by checking the correctness of additional triads. Both the process and its results are described in Section~\ref{cucode:section:permutation}.

In combination with executing known triads directly from \ac{ROM} and extracting their side effects, we can correlate the emulated instructions with their counterparts with known addresses.

\subsection{Microcode Emulation}
\label{cucode:section:re:emulation}
In order to gather a sufficiently large number of data points to
reverse engineer the fine grained mapping of the \ac{ROM} addresses,
we implemented a microcode emulation engine.
This emulation engine is designed to replicate the behavior of the \ac{CPU} during the execution of a given triad.
This means that for any given input,
the output of both the physical \ac{CPU} and our emulation engine should be identical.
As our analysis framework is implemented in Python,
we also chose this language to implement the emulator.
The emulator is designed to interpret the bitstrings extracted from
the \ac{CPU} and first disassembles them using our framework.
For each individual micro-op, this yields the operation as well as the source and target operands.
The operations itself are implemented as Python lambdas modifying the indicated registers.
This allows for simple extension of the supported instruction list.
For each triad the emulator returns a changeset indicating the changed registers and their new values.
Currently this is done on a triad-by-triad basis to support our reverse engineering method.
However, by supplying the changed register set as the input state for the next
triad, the emulation can be performed for any number of triads in sequence.
The emulation engine currently supports all of the identified arithmetic microcode operations.
Additionally, we supply a whitelist of instructions that produce no visible effect on the specified registers.
While these instructions have side effects when executed on the \ac{CPU},
they are treated as no-ops,
because only the visible state of the registers is considered in our further analysis.
The instructions and their behavior are based on previous reverse engineering results.
We ensured that we correctly identified a certain instruction by executing the bitstring of the instruction
in a microcode update applied to a real CPU and observing the effects on the specified registers with varying inputs.

However,
as the \ac{ROM} contains operations that implement unknown behavior,
most importantly reading and writing internal status registers
or collecting information on the currently executed instruction,
we were unable to accurately emulate all of the triads.
Also the readout itself introduced both potential bit errors as well as sections that
are unable to be read due to dust particles or other disturbances in the raw image.
We thus opted to only consider triads for further analysis that (i) contain
only known instructions and (ii) were not part of an unreadable section.
This emulation yielded the behavior of triads with known physical addresses for a given input state.
The input state assigned a different value to every x86 and usable microcode register.
During testing we observed that not all microcode registers can be freely assigned to,
some will trigger erratic \ac{CPU} behavior leading to crashes or loss of control.
Thus,
we had to exclude certain registers from our tests.
Our input and output state contains all six x86 general purpose registers (we excluded the stack
manipulation registers EBP and ESP) as well as in total 22 internal microcode registers.

To gather the behavior for known logical addresses,
we forced execution of each \ac{ROM} triad directly on the \ac{CPU}.
For this execution,
we chose the same input state that was previously used for the emulation.
The input state was set by a sequence of x86 instructions setting the x86 registers to the chosen values.
The microcode registers were then set after entering microcode by a sequence
of micro-ops preceding the jump to the triad address to be tested.
The output was gathered by writing out the changed registers as specified
by our emulator to x86 registers using microcode executed after the tested triad.
Due to the different values for each register,
we could determine which register was used as an input in the tested triad as well as the operation performed on it.
However,
we also had to exclude a large number of logical addresses as those triads lead to a
loss of control or showed a behavior that was independent of the given input state.
In combination,
these two tests yielded a collection of address pairs consisting out of the
physical address of a candidate triad and the logical address of the triad.

{\captionsetup[figure]{skip=5pt}
\begin{figure*}[!t]
	\begin{minipage}{2\columnwidth}
	\centering
	\resizebox{0.80\linewidth}{!}{
		\includegraphics{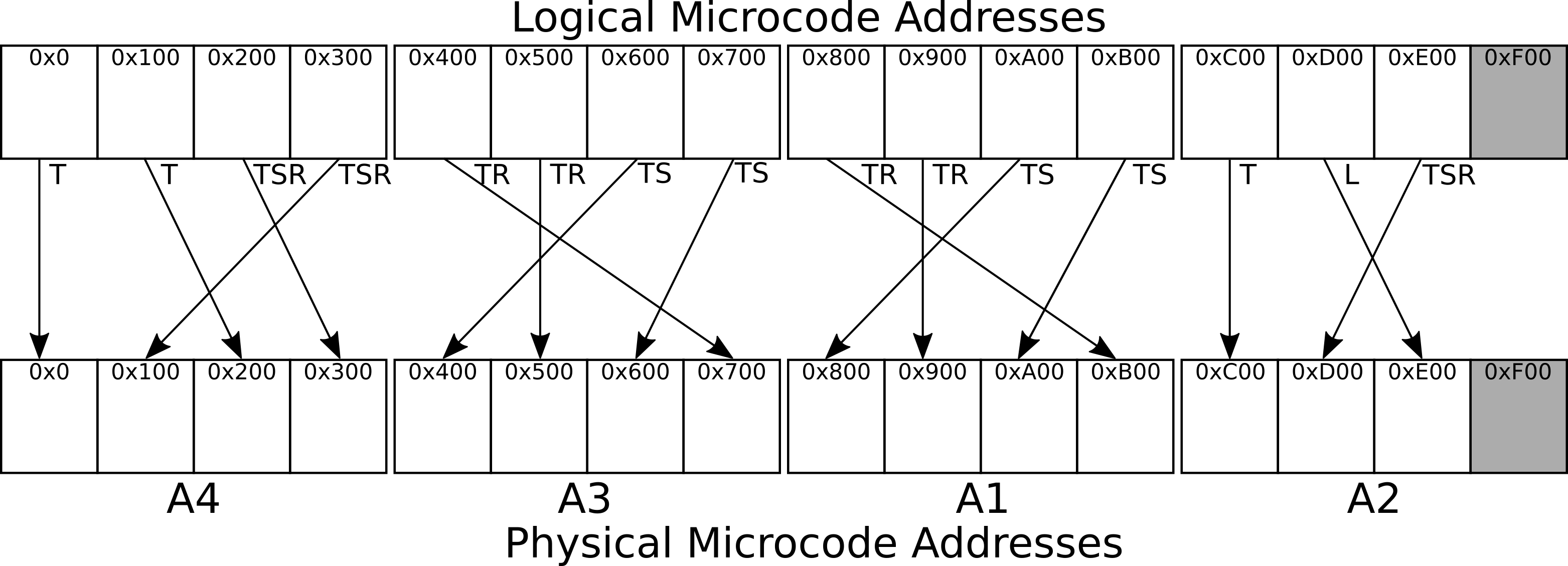}
	}
	\caption{Translation of logical to physical microcode \ac{ROM} addresses.}
	\label{cucode:figure:rom:mapping}
	\end{minipage}
\end{figure*}
}

\subsection{Permutation Algorithms}
\label{cucode:section:permutation}

After gathering the microcode address pairs, we had to reconstruct the function used to map these onto each other.
Due to the hardware layout and hardware design possibilities, we determined a number of different candidate permutation functions.
Additionally, we used the data points gathered in the previous step to develop new algorithmic options.
We then applied these possible functions in combination to test whether they were used for a specific triad.

Via this empirical testing, we found that the \ac{ROM} uses the following permutations:
\begin{itemize}
	\item T: table based 16 triad-wise permutation, illustrated in Table~\ref{cucode:figure:algorithms:table}
	\item R: reverse counting direction, mapping higher physical address triads to lower logical addresses
	\item S: pairwise swap two consecutive triads
	\item L: custom table based 16 triad-wise permutation for last block, illustrated in Table~\ref{cucode:figure:algorithms:table}
\end{itemize}

To determine the combination of permutations used for a specific address pair,
we verified the possibilities by calculating the physical address for the given logical address.
If the result matches the expected value,
the combination is correct.
The found combination is then used to calculate the physical addresses for the rest of the data points.
Once a mismatch is found, the first approach is repeated to determine the next combination of permutations.

We determined that the mapping function is constant for 256 triads at a time,
then the combination of algorithms changes.
We also had to account for potentially swapped 256 triad blocks,
so in case of a mismatch the remaining triad blocks in a region were then considered.
This yielded the mapping algorithm for all but the last 256 triads.
The last block uses a different mapping algorithm that was reconstructed manually.
The detailed mapping of all triad blocks is given in Figure~\ref{cucode:figure:rom:mapping}; Table~\ref{cucode:figure:algorithms:table} illustrates the permutation algorithms T and L.

\begin{table}[!htb]
\centering
\begin{tabular}{|l|l|l|}
\hline
Physical & logical - T & logical - L\\
\hline
\texttt{0x00} & \texttt{0x00} & \texttt{0x00}\\
\texttt{0x10} & \texttt{0x20} & \texttt{0x10}\\
\texttt{0x20} & \texttt{0x40} & \texttt{0x20}\\
\texttt{0x30} & \texttt{0x60} & \texttt{0x30}\\
\texttt{0x40} & \texttt{0x80} & \texttt{0x40}\\
\texttt{0x50} & \texttt{0xA0} & \texttt{0x50}\\
\texttt{0x60} & \texttt{0xC0} & \texttt{0x60}\\
\texttt{0x70} & \texttt{0xE0} & \texttt{0x70}\\
\texttt{0x80} & \texttt{0x10} & \texttt{0xF0} (RS)\\
\texttt{0x90} & \texttt{0x30} & \texttt{0xE0} (RS)\\
\texttt{0xA0} & \texttt{0x50} & \texttt{0xD0} (RS)\\
\texttt{0xB0} & \texttt{0x70} & \texttt{0xC0} (RS)\\
\texttt{0xC0} & \texttt{0x90} & \texttt{0xB0} (RS)\\
\texttt{0xD0} & \texttt{0xB0} & \texttt{0xA0} (RS)\\
\texttt{0xE0} & \texttt{0xD0} & \texttt{0x90} (RS)\\
\texttt{0xF0} & \texttt{0xF0} & \texttt{0x80} (RS)\\
\hline
\end{tabular}
    \caption{Translation of addresses for the T and L algorithms. The L algorithm applies the R and S permutations to the higher addresses after the table based permutation.}
    \label{cucode:figure:algorithms:table}
\end{table}

%% file: sections/primitives.tex
\section{Microcode Primitives}
\label{cucode:section:primitives}

Microcode programs supported by modern processors combined with the ability to
update this microcode can provide a range of useful security primitives that can be used to build system defenses. 
In the following, we explore several key primitives and discuss in
Section~\ref{cucode:section:case_study} how system defenses can be implemented based on
our analysis results described in the previous section. 

\par{\bf Enabling or disabling CPU features at runtime}

Despite recently uncovered security issues such as \textsc{Spectre} and \textsc{Meltdown}~\cite{Lipp2018meltdown,projectzeromeltdownspectre,Kocher2018spectre},
speculative execution is an important feature that enables the performance of current \ac{CPU} families.
While the na\"ive coun\-ter\-mea\-sure---disabling speculative execution completely---provides a high level of security,
it significantly reduces the performance of a given system.
However, if the speculative execution could be disabled only temporally or only for certain program states,
a trade-off between security and performance could be implemented.

Another example of a feature that can be used by both benign and malicious applications is the availability of high-resolution timers.
Such timers allow an attacker to abuse microarchitectural timing side channels to gather information from otherwise inaccessible contexts~\cite{crypto:1996:kocher,Hund,pakt:2012,206170}.
In both cases, microcode can improve security by applying a fine-grained permission model on top of existing protection mechanisms
by restricting features to certain applications or contexts only.

\par{\bf Intercepting low-level \ac{CPU} processes} 
A core functionality of microcode is the decoding of instructions.
By intercepting this step during the execution of x86 code, it is possible to apply fine-grained control over the behavior of
instructions, programs, and the system as a whole.
From a security perspective, additional functionality can be added to existing instructions,
special handling for corner cases can be inserted, and security checks can be included.

Besides changing and extending the instruction decoding,
it is also possible to influence other aspects of the \ac{CPU}'s operation.
For example, the exception handling mechanism is implemented with the help of microcode.
Before an exception reaches the kernel-level x86 code, microcode can change the metadata passed to the kernel or handle the
exception without involving the kernel at all.
By directly modifying the exception handling in microcode, expensive context switches can be avoided.
This allows, for example,
special handling of page faults to implement page-based memory separation in a way that is completely transparent to the kernel.

\par{\bf Isolated execution environment} 
The microcode engine provides a fully-featured execution environment that cannot be
intercepted by the running kernel in any way.
Any exception delivered while microcode is running will be stalled until the current decoding is complete.
Moreover, any state that is not explicitly written out will be contained in the microcode engine and cannot be accessed.
More specifically, both the running kernel and hypervisors are unable to inspect the execution state of the microcode engine.
This provides an enclave-like environment in which computations on sensitive data can be performed in an opaque way.
Only the results will be passed to the operating system,
protecting secret keys or other data inside the microcode.

\par{\bf Extending and modifying the x86 instruction set} 
By either reusing no longer used x86 instructions or adding entirely new
instructions to the decoding process,
microcode can enable functionality not found in the standard x86 instruction set architecture.
These instructions can for example implement more complex semantics that are tailored to a specific use case.
By condensing calculations into fewer instructions, caches are utilized more effectively,
increasing performance.
Besides performance improvements, new primitives can be added with new instructions.
As microcode can change the access level between operations,
it is able read and write kernel-only data structures.
Combining this with fine-grained checks enables fast access to otherwise privileged functions, without support of the running
kernel.

\section{Case Studies of Microcode Defenses}
\label{cucode:section:case_study}

Based on the security primitives discussed above, we now present designs and \textit{proof-of-concept} implementations of our microcode-assisted systems defenses and constructive microcode applications. For each case study, we first briefly motivate the primitive, present the design and implementation, and conclude with an evaluation and discussion of advantages and drawbacks of our approach. Based on these case studies, we demonstrate that microcode indeed improves properties of those applications with regards to performance, security, and complexity.
The microcode programs and supporting infrastructure are publicly available~\cite{microcode:amd_microprograms}.

The current state of the programs does not feature a mechanism for runtime configuration,
however this is can be achieved in different ways.
As it is possible to load microcode updates during runtime,
the operating system can apply an update to enable or disable certain features.
It is also possible to add dedicated flags in the thread or process control structures created by
the operating system to signal which features should be enabled for a certain thread.
However, both approaches require support from the OS to either synchronize the microcode update procedure
across all \ac{CPU} cores or allocate and initialize the configuration fields for every new thread.
Another option is to use processor-internal storage to store configuration variables.
Tests showed that a writable region of memory exists that can be used to store these variables.
Unfortunately, further experiments are needed to ascertain the nature
of this memory region and the side effects of changing it at runtime.

We evaluate the performance of our case studies with microbenchmarks of the affected instructions. To this end, we determine the execution time in cycles as measured via the \texttt{rdtsc} instruction.
It provides the value of the \ac{TSC},
a counter which is incremented each clock cycle.
The used code snippet for performance benchmarks is illustrated in Figure~\ref{ucode:figure:measurement}. All tests were performed on an AMD Sempron 3100+ running the minimal operating system developed by Koppe~\etal~\cite{koppe2017reverse}. In the following, the cycle counts are given without the overhead of the measurement setup itself, which adds 65 cycles to every execution.
Further improvements to the performance properties of the defenses are possible with a greater understanding of the underlying hardware. This requires either more work on reverse engineering more details, especially in regards to scheduling, or, to fully utilize the existing hardware, assistance of the \ac{CPU} vendors.

{\lstset{language=[x86masm]Assembler, deletekeywords={size}}
\begin{figure}
\begin{lstlisting}
xor eax, eax
xor edi, edi
cpuid
rdtsc
xchg edi, eax

; benchmarked instruction
shrd ebp, ecx, 4

cpuid
rdtsc
sub eax, edi
\end{lstlisting}
\caption{Microbenchmark setup to determine the execution time in cycles of \texttt{shrd} (double precision right shift). The modern \texttt{rdtscp} instruction variant is not available on the tested K8 CPU, thus the \texttt{cpuid} instruction is used to serialize the instruction execution.}
\label{ucode:figure:measurement}
\end{figure}
}

\subsection{Customizable RDTSC Precision}
\label{cucode:section:case_study:rdtsc}

\par{\bf Motivation.}
Previous works demonstrated the possibility to reconstruct the memory
layout~\cite{Hund,gras2017aslr,oren2015spy} using timing side channels.
More recently the \textsc{Spectre} and \textsc{Meltdown} attacks have shown in a spectacular
way~\cite{Kocher2018spectre,projectzeromeltdownspectre,Lipp2018meltdown} that it is possible
to break the fundamental guarantees of memory isolation on modern systems.
A common aspect of these attacks is the usage of high-resolution timers to observe the timing side channels.
Due to these dangers,
modern browsers limit the accuracy of high-resolution timers to a recommended value~\cite{w3c-rec-candidate}.
While this does not eliminate all timing sources~\cite{schwarz2017fantastic,kohlbrenner2016trusted},
it raises implementation complexity of attacks and provides a mitigation against common exploits.

On the native level the timing information is commonly queried using the \texttt{rdtsc} instruction.
The x86 architecture allows limiting \texttt{rdtsc} to kernel space only.
Any attempt of executing this instruction from user space will lead to a fault.
Building upon this fact,
the operating system can limit the resolution of the timer available to user programs.
Upon receiving the corresponding fault, the operating system queries the \ac{TSC} itself,
reduces the resolution accordingly, and passes the timestamp onto the program.
Note that this incurs a significant performance overhead due to the necessary context switches.

\par{\bf Design and Implementation.}
Since we are able to change x86 microcode behavior,
our goal is to implement a functionality similar to the
browser mitigation for the native \texttt{rdtsc} instruction.
In addition,
our solution should be able to reduce the accuracy to a pre-defined value
without incurring unnecessary overhead in form of context switches.
To this end,
we intercept the execution of \texttt{rdtsc} and before the \ac{TSC} value is made available to the application,
we set a pre-defined number of lower bits to zero.
Note that the amount of zeroed bits is configurable (in the microcode
update) to provide a trade-off between accuracy and security.

\par{\bf Evaluation and Discussion.}
While the default implementation of \texttt{rdtsc} takes 7 cycles to execute,
our custom implementation takes a total of 15 cycles to complete.
This overhead is due to the switch to microcode \ac{RAM} and the additional
logical \textsf{AND} operation to clear the lower bits of the \ac{TSC} value.
The \ac{RTL} representation of our \texttt{rdtsc}
implementation is shown in the appendix in
Listing~\ref{cucode:listing:rdtsc}.

Even though our solution doubles the execution time, %
it is far faster than the approach where the kernel needs to trap the raised interrupt.
At the same time,
our security guarantees are comparable to the discussed browser mitigations.
While raising the bar,
timing attacks are still possible by using methods described by
Schwarz~\etal~\cite{schwarz2017fantastic} and
Kohlbrenner~\etal~\cite{kohlbrenner2016trusted}.

\subsection{Microcode-Assisted Address Sanitizer}
\label{cucode:section:case_study:hwasan}
\par{\bf Motivation.}
\ac{ASAN}~\cite{ASAN} is a compile-time instrumentation framework that introduces checks for every memory access in order to uncover both spatial and temporal software vulnerabilities. In particular, temporal faults such as \textit{use-after-free} bugs present an important class of memory corruption vulnerabilities that have been used to exploit browsers and other software systems~\cite{van2017dynamics}.
\ac{ASAN} tracks program memory state in a so-called \textit{shadow
map} that indicates whether or not a memory address is valid.
Therefore,
\ac{ASAN} inserts new instructions during compilation to perform the
checks as well as an instrumentation of allocators and deallocators.
In addition,
\ac{ASAN} enforces a quarantine period for memory regions and thus prevents them from being re-used directly.
However,
this instrumentation incurs a performance overhead of roughly 100\%.

To overcome the performance penalty and reduce the code size,
the authors of \ac{ASAN} also discussed how a hardware-assisted version, dubbed \ac{HWASAN}, could theoretically be implemented~\cite{HWASAN}. The basic idea is to introduce a new processor instruction that performs access checks.
The general principle of the new instruction is illustrated in Figure~\ref{ucode:figure:algorithms:hwasan}.
It receives two parameters: the pointer to be accessed and the memory access size.
The instruction then validates the memory access and its size with the help of the shadow map.

{\lstset{language=C}
\begin{figure}
\begin{lstlisting}
CheckAddressAndCrashIfBad(Addr, kSize) {
	ShadowAddr = (Addr >> 3) + kOffset;
	if (kSize < 8) {
		Shadow = LoadByte(ShadowAddr);
		if (Shadow && Shadow <= (Addr & 7) + kSize - 1)
			ReportBug(Addr);
	} else {
		Shadow = LoadNBytes(ShadowAddr, kSize / 8);
		if (Shadow)
			ReportBug(Addr);
	}
}
\end{lstlisting}
   \caption{Pseudocode of the \ac{HWASAN} instruction~\cite{HWASAN}; \texttt{kSize} is the size of the memory access and \texttt{kOffset} is a compile time constant that specifies the location of the shadow map.}
\label{ucode:figure:algorithms:hwasan}
\end{figure}
}

\par{\bf Design and Implementation.}
Instead of requiring a hardware change to add the new \ac{HWASAN} instruction,
we design a scheme to implement \ac{HWASAN} in microcode.
Similarly to Figure~\ref{ucode:figure:algorithms:hwasan},
we perform the checks accordingly and raise a fault in case an invalid memory access is detected.
To provide a clear separation between application code and instrumentation,
we implemented the checking in a single instruction.
For practical reasons,
the interface should be easy to add to existing toolchains.

In our implementation,
we chose to reuse an existing but unused x86 instruction, in this case the instruction \texttt{bound}.
Since the check requires address and size of the memory access,
we changed the interface of this instruction in the following way:
the first operand indicates the address to be accessed,
while the second operand indicates the access size.
We want to emphasize that that our microcode instrumentation can be emitted
without changes to an existing x86 assembler using the following syntax:
{\lstset{language=[x86masm]Assembler, deletekeywords={size}}
\begin{lstlisting}
bound reg, [size]
\end{lstlisting}
}

Similarly to \ac{ASAN},
our instruction is inserted in front of every memory access during compilation.
We also use the same shadow map mechanism and base address,
hence the instrumentation requires no additional changes.
However,
the key difference is the compactness and that no externally visible state is changed.
In case the memory access is valid,
the instruction behaves as a \texttt{nop},
but if an invalid access is passed, a defined action is taken.
To this end, our prototype implementation currently support three methods of error reporting:
\begin{enumerate}
      \item raising a standard access violation,
      \item raising the \texttt{bound} interrupt, and
      \item calling a predetermined x86 routine.
\end{enumerate}
Note that the first two options rely on the availability of an exception handling mechanism,
while the latter option is self-contained and works even without kernel support.

\par{\bf Evaluation and Discussion.}
While the checking algorithm is semantically the same,
we observed a performance advantage of our solution.
The default \ac{ASAN} implementation for a (valid) 4 byte load requires 129 cycles to complete,
our version requires only 106 cycles.
Another advantage of our implementation is that no x86 register is changed during its execution:
instead of using x86 general purpose registers, our implementation
stores temporary values in ephemeral microcode-internal registers.
This means the insertion of the instrumentation does not increase the register
pressure and does not cause additional register spills to the stack.
This is in comparison to the original \ac{ASAN} implementation
which uses two additional x86 registers to hold temporary values.
The overhead of additional register spills is not included in
our benchmark as it is highly dependent on the surrounding code.
The \ac{RTL} representation of our \ac{HWASAN} implementation
can be found in our Github repository~\cite{microcode:amd_microprograms}.

\subsection{Microcoded Instruction Set Randomization}

\par{\bf Motivation.}
In order to counter so-called \textit{code-injection} attacks,
a series of works investigated \ac{ISR}
schemes~\cite{sovarel2005s,papadogiannakis2013asist,portokalidis2010fast,hu2006secure,barrantes2003randomized,kc2003countering}
with the goal of preventing the correct execution of maliciously injected code.
To this end,
the instruction encoding is randomized (e.g.,
using an \textsf{XOR} with a pre-defined key) for all or a subset of instructions,
so that the adversary does not know the semantics of a randomized instruction.
Note that recently published advanced schemes also aim to mitigate
\textit{code-reuse} attacks using strong cryptographic encryption
algorithms~\cite{sinha2017reviving}.
However, most schemes require hardware support,
which prevents their deployment to \ac{COTS} \acp{CPU}.

\par{\bf Design and Implementation.}
Our \ac{ISR} scheme removes the link between the actual x86 operation and its semantics,
and thus an adversary is unable to infer the meaning of an instruction
stream even if disassembled during a \ac{JIT}-\ac{ROP} attack.
In order to be robust even when facing code-reuse or \ac{JIT}-\ac{ROP} attacks,
we assume fine-grained code randomization or software diversification.

Our proof-of-concept implementation supports six different operations:
memory load, register move, add,
left and right shift,
and exclusive or.
Each operation can be freely assigned to any microcoded x86 instruction
that allows for one register operand and one memory operand.
This assignment effectively binds the executed x86 code to a specific instance of the \ac{ISR}.
Execution is only possible
if the semantics implemented in microcode for each instruction match the one used when generating the x86 code.
Note that due to this varying assignment and the variable instruction length of the affected opcodes,
it is not possible to assemble a \ac{ROP} chain or shellcode matching all possibilities.
Additionally, we support masking of input and output values before they
are written to or read from potentially attacker-accessible storage,
including system memory and registers.

To facilitate the translation of existing x86 code to opcodes using the newly introduced semantics of the \ac{ISR},
we implemented a \textit{transpiler}.
This transpiler processes a stream of disassembled x86 instructions and replaces all
occurrences of supported opcodes with the appropriate opcodes with changed semantics.
The selection of the replacement opcode is performed based on the assignment in the corresponding microcode update.
The input to the transpiler is thus the source instruction stream and the
mapping of x86 instructions to semantics as implemented by the \ac{ISR},
the output is a modified instruction stream.
This output stream can them be assembled by a standard x86 assembler,
as no new instructions are introduced.

\par{\bf Evaluation and Discussion.}
We evaluate the performance of our implementation by comparing the runtime
(measured in cycles according to the test setup described previously)
of a toy example consisting only out of supported opcodes with the corresponding transpiled version.
Our measurements indicate that our microcoded \ac{ISR} scheme introduces
an overhead of 2.5 times on average over a set of 5 different examples,
compared to the same code running natively.
This overhead is mainly due to replacing non-microcoded instructions (that normally
take 1-3 cycles) with microcoded instructions that require at least 7 cycles,
including the additional overhead of switching to microcode \ac{RAM} execution.
We provide one of the test cases in Listing~\ref{cucode:listing:isr} in the appendix.
Note that the cumulative performance of instruction streams may vary due to pipelining and parallel execution.
This is especially visible if instructions covered by the \ac{ISR} are mixed with standard x86 instructions.
As our toy examples exclusively use transpiled instructions,
we arrive at the worst case overhead.
Since the \ac{ISR} can implement more complex semantics such as a multiply-accumulate,
the cycle overhead can be reduced with a more advanced transpiler.
We want to emphasize that our \ac{ISR} does not require hardware changes compared
to previous schemes and thus can be deployed on \ac{COTS} \acp{CPU} with a microcode update.

\subsection{Microcode-Assisted Instrumentation}

\par{\bf Motivation.}
Traditional binary defenses often suffer from either significant performance overhead or incompleteness.
This is typically due to the reliance on dynamic instrumentation or static binary rewriting.
However,
with the ability to change the behavior of x86 instructions via a microcode update,
it is possible to intercept only specific instructions without impacting performance of unrelated code.
Hence,
a microcode-assisted instrumentation leverages synergies of minimal performance overheads
of static binary rewriting and completeness of dynamic instrumentation solutions.

{\captionsetup[figure]{skip=5pt}
\begin{figure}[!t]
	\centering
	\resizebox{0.65\linewidth}{!}{
		\includegraphics{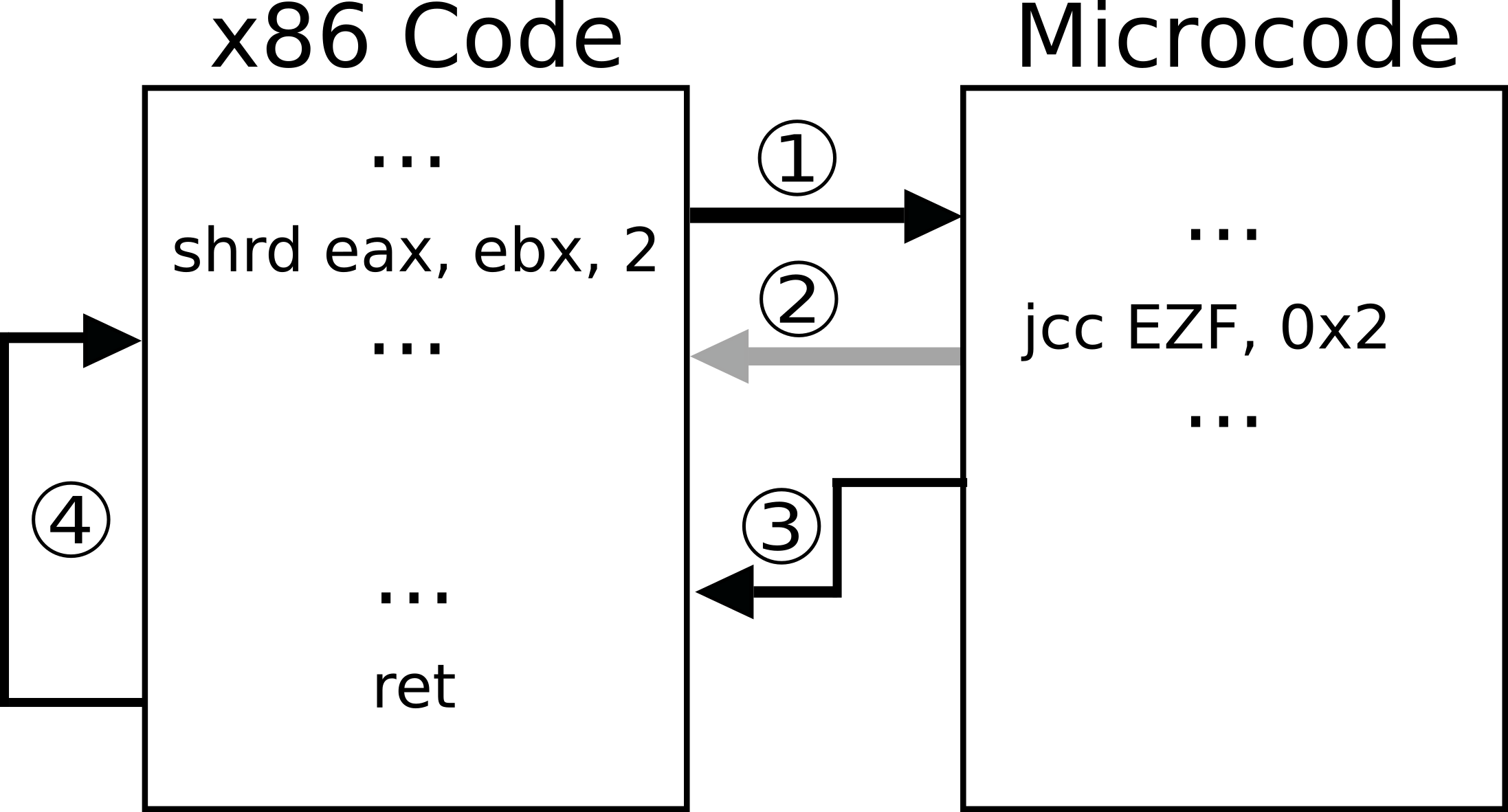}
	}
	\caption{Control flow of an instrumentation.}
	\label{ucode:figure:primitives:instrumentation}
\end{figure}
}

\par{\bf Design and Implementation.}
We designed a microcode-as\-sist\-ed instrumentation scheme that allows generation of microcode updates that
intercept a specific instruction and upon execution of this instruction, the control is transferred to a specific address.
This address contains standard x86 code to perform the instrumentation and finally resume execution.
The microcode update can additionally contain a custom-tailored filtering,
so that the x86 handler is only invoked on specific conditions.
As the filtering is implemented directly in microcode,
the overhead of changing the x86 execution path which can invalidate
branch prediction and caches is only occurred when needed.

\par{\bf Evaluation and Discussion.}
To test the viability of the instrumentation,
we implemented a proof-of-concept microprogram that instruments \texttt{shrd} to call
an x86 handler if a certain constant is detected in the argument register.
The control flow is illustrated in Figure~\ref{ucode:figure:primitives:instrumentation}.
Upon execution of the instruction,
\raisebox{.5pt}{\textcircled{\raisebox{-.9pt} {1}}}~control is transferred to the microcode \ac{RAM}.
\raisebox{.5pt}{\textcircled{\raisebox{-.9pt} {2}}}~As a filter,
we check if the argument register is equal to a constant.
In case the filter does not match,
the instruction is executed normally and x86 execution continues after \texttt{shrd}.
In case the filter matches,
\raisebox{.5pt}{\textcircled{\raisebox{-.9pt} {3}}}~the current instruction pointer
is pushed onto the stack and the x86 instrumentation part gains control,
comparable to a \texttt{call} instruction in x86.
Once our instrumentation gains control, it can perform any number of calculations
and is not constrained by the size limitations of the microcode \ac{RAM}.
~\raisebox{.5pt}{\textcircled{\raisebox{-.9pt} {4}}}~Finally,
the instrumentation continues the normal execution by returning to the interrupted code.

We also conducted a performance benchmark to determine the overhead introduced by our instrumentation
for the case where the microcoded condition does not hold --- illustrated with~\raisebox{.5pt}{\textcircled{\raisebox{-.9pt} {2}}}~in Figure~\ref{ucode:figure:primitives:instrumentation}.
In this case, the x86 execution should continue as fast as possible
in order to reduce the overhead for any code not to be inspected.
We use the \texttt{shrd} instrumentation for this test and measure the performance according to the described test setup.
The original implementation of \texttt{shrd} executed in 2 cycles,
our test case took 8 cycles.
This overhead is mainly due to the switch to microcode \ac{RAM}
and the two triads inserted for the instrumentation check.
The microcode \ac{RTL} of the \texttt{shrd} instrumentation is available in our Github repository~\cite{microcode:amd_microprograms}.

While the execution time of the single instruction is increased substantively,
this overhead is fixed for any semantic the instruction originally implements.
This implies that our instrumentation only adds 6 cycles to perform its own check,
regardless of the original run time of the instruction.
Additionally, we do not introduce a conditional x86 branch,
which further increases the overhead due to potential branch mis-predictions.
Moreover, our implementation does not use scratch x86 registers and thus does
not increase register pressure or causes additional memory accesses.
Finally, the overhead is only introduced for instructions that are to be inspected,
the rest of the execution is not impacted.
This is in contrast to existing dynamic instrumentation frameworks,
such as Valgrind~\cite{nethercote2007valgrind}, PIN~\cite{luk2005pin} or DynamoRIO~\cite{dynamorio},
which increase the execution time for all instructions. For a lightweight instrumentation, the overheads induced by these tools are about 8.3, 2.5 or 5.1 times, respectively~\cite{luk2005pin}.

On top of our framework,
any binary instrumentation relying on intercepting of a \textit{small} number x86 instructions can be realized.
Note that a current limitation is that only microcoded instructions can be intercepted, however,
this is a limitation of the current reverse engineering progress.
Previous work indicated the possibility of intercepting all instructions,
included non-microcoded ones.

\subsection{Authenticated Microcode Updates}
\label{cucode:section:microcode_update}

\par{\bf Motivation.}
While the insufficiently protected microcode update mechanism of
AMD K8 and K10 processors enabled the research in the first place,
it simultaneously poses a major security issue:
an attacker can apply any update of her choosing,
which was demonstrated by Koppe~\etal~\cite{koppe2017reverse} by developing stealthy microcoded Trojans.
However, as the microcode update mechanism itself is implemented in microcode,
it is possible to develop a protection mechanism in the form of a microcode update that can provide limited security guarantees.
We implement a proof-of-concept that demonstrates the feasibility of such a scheme on the affected \acp{CPU}.

\par{\bf Design and Implementation.}
In order to mitigate the risk associated with the current scheme,
a microcode update mechanism is required that only accepts authenticated updates.
However,
given the ephemeral nature of microcode updates,
this countermeasure requires either a hardware re-design or a trusted application (e.g.,
a part of \ac{UEFI} with secure boot) that applies a suitable microcode update early during boot.
In particular,
this update must then verify each further update attempt using proper cryptographic primitives.
At the same time,
due to the limited space in the microcode update,
the verification has to be small in terms of code size.
Note that performance is of lesser priority in this case since
microcode updates are typically only performed once per system start.

Our implementation extends the \texttt{wrmsr} instruction, which is used to start the microcode update, to enforce the following properties for the microcode update:
\begin{enumerate}
   \item The update includes 32 triads, the maximum possible number on the K8 architecture. The vendor-supplied updates are always padded to this length.
   \item A \ac{HMAC} is appended to the update directly after the last triad.
   \item The \ac{HMAC} is correct for the full update, including the header. The inclusion of the header in the authenticated part protects the match registers and thus the affected instructions. The key of the \ac{HMAC} is included in the initial microcode update.
\end{enumerate}

For our implementation,
we choose the block cipher \ac{TEA}~\cite{wheeler1994tea} due to the simplicity of
its round function which results in a small code size in the microcode \ac{RAM}.
This is especially important as our current understanding of microcode semantics
only allows loading of 16-bit immediate values per microcode operation.
Hence,
loading of a single 64-bit constant requires a total of 8 operations or nearly
three triads (note that the whole microcode update is limited to 32 triads only).
While it would be preferable to implement a strong cryptographic algorithm such as \ac{AES},
these commonly require S-Boxes,
which we cannot support due to code size constraints.

\par{\bf Evaluation and Discussion.}
As we extend the standard update mechanism with an additional verification of the entire microcode update,
we incur a significant performance hit.
In our tests, applying a maximum length update takes 5,377 cycles without the authenticated update mechanism.
With our deployed authentication scheme, loading the same update requires 68,525 cycles.
This increase is expected due to the added verification.
As the update is only applied once during system boot, the performance hit is still negligible.
For comparison,
the AMD 15h architecture (Bulldozer etc.) requires 753,913 cycles on average for an update~\cite{tr:2014:chen}.
This generation likely uses a public key scheme to verify the update.

Due to code size limitation we were limited to the simple and small \ac{TEA} algorithm and could not implement a public key verification scheme.
However, if the update authentication mechanism were contained in the microcode \ac{ROM} directly,
the code size would not be as restricted.
While our \ac{ROM} readout indicates a very high usage of the available triads,
there are still more padding triads present than would fit into a microcode update.
In our prototype implementation, the user can decide which updates to trust,
or given the possibility to disassemble the updates,
even which parts of an update should be applied.
This allows for a finer control over the hardware than what would be possible using only a vendor-accessible signature
method.
The \ac{RTL} of our microcode authentication scheme is available in our Github repository~\cite{microcode:amd_microprograms}.

\subsection{$\mu$Enclave}

\par{\bf Motivation.}
Intel \ac{SGX}~\cite{iacr:2016:86} is an instruction set extension that introduces
the creation of isolated, trusted execution environments with private memory regions.
These so-called \textit{enclaves} are protected from processes even at
high privilege levels and enable secure remote computation.
Inspired by \ac{SGX} we designed and implemented a \textit{proof-of-concept} enclave functionality,
dubbed \textit{$\mu$Enclave}.
$\mu$Enclave can remotely attest that code indeed runs inside the enclave and ensures confidentiality of data.
We can thus retrofit basic enclave functionality to older \acp{CPU} not offering a comparable solution.
Additionally, we use this case study to illustrate the isolation property of microcode.

\par{\bf Design and Implementation.}
We leverage the separate microcode \ac{IDU} to establish an isolated execution environment.
The other decode units are halted while the microcode \ac{IDU} is active by design of the microarchitecture.
Due to these isolation properties we can safely assume that x86 code,
even when running with kernel-level privileges,
cannot interfere with the enclave program implemented in microcode at run time.

$\mu$Enclave is based on the authenticated microcode update mechanism,
presented in Section~\ref{cucode:section:microcode_update},
and the following strategy:
\begin{enumerate}
      \item The trust is built upon the symmetric key contained in the first microcode update applied early during boot by \ac{UEFI}. The entity controlling that key may be a chip manufacturer, software vendor, or the end-user. The entity has to ensure that payload microcode updates contain only benign behavior before signing it. 
\item The program that is supposed to run in the $\mu$Enclave is implemented in microcode and embedded in a signed payload microcode update.
\item The enclave program may perform arbitrary computations and access virtual memory. The enclave program may write sensitive data into \ac{RAM}, but it must ensure security properties like authenticity, integrity, and secrecy itself using signing and encryption.
\item The enclave program can remotely attest that it indeed runs within the enclave by signing a message with the symmetric enclave key.
\end{enumerate}

\par{\bf Discussion.}
In combination with a challenge-response protocol,
$\mu$Enclave enables remote attestation and additional services of the enclave
can be exposed either via augmenting x86 instructions or adding new \acp{MSR}.
A major drawback of $\mu$Enclave is the restricted code size due to the microcode \ac{RAM} size.
This limitation can be lifted by either implementing a small virtual machine and interpreting signed bytecode from
main memory or iteratively streaming signed microcode from main memory to microcode \ac{RAM} as it executes.
For the latter, we are missing the micro-ops that can write to microcode \ac{RAM}.
While our current implementation does not support either approach,
this is not a fundamental limitation of $\mu$Enclave.

When compared to sophisticated trusted execution environments such as Intel \ac{SGX} or ARM TrustZone,
$\mu$Enclave is more cumbersome to use.
As the enclave code needs to be written as microcode,
the development requires experience with this environment.
Additionally, the limited code size limits the selection of
cryptographic primitives to those with very small implementations.
This results in the use of less secure cryptographic algorithms and thus lower security guarantees.
Finally, the \ac{CPU} lacks hardware support and acceleration for cryptographic operations.
This means, for example,
that the attestation needs to be implemented by the programmers of enclave code themselves.
However, $\mu$Enclave can be used on older \acp{CPU}
that do not provide the mentioned vendor supplied solutions.
As such, it is possible to add similar primitives to legacy \acp{CPU} without requiring a hardware change.

%% file: sections/discussion.tex
\section{Discussion and Future Work}
\label{cucode:section:discussion}

In this section, we discuss benefits and challenges of microcode-assisted system defenses and review limitations of microcode in general and of our reverse engineering approach in particular. Furthermore, we present and discuss potential topics for future work such as microcode-assisted shadow stacks, lightweight syscalls as well as information isolation. We also shed light on how microcode Trojans can be detected. 

\subsection{Microcode for System Defenses}
\label{cucode:section:discussion:defenses}

Modern processor microcode and the ability to update microcode can provide useful primitives such as enabling or disabling \ac{CPU} features at runtime, intercepting instruction decoding or other microarchitectural processes to modify existing behavior, providing a small execution environment isolated from the operating system kernel, and bypassing some boundaries of the x86 \ac{ISA} to implement new features. We have shown in Section~\ref{cucode:section:case_study} that these primitives enable the implementation of some defensive schemes like customizable accuracy of the built-in x86 timer and $\mu$Enclave in the first place. Other defenses such as microcoded \ac{HWASAN} and \ac{ISR} benefit from these primitives with regard to performance overhead and complexity. With more knowledge about microcode, additional defenses like opaque shadow stacks and information isolation can be built, as we discuss in Sections~\ref{cucode:section:discussion:shadowstacks} and Section~\ref{cucode:section:syscalls}. However, the generality of microcoded primitives suffers due to the limited number of processor models that currently accept custom microcode updates. We argue that the introduction of an open and documented microcode API could benefit system security research and future defensive systems. Such an API has to address several challenges like abstracting the underlying changes through processor generations, conflict handling for concurrent updates, and ensuring system stability. In order to avoid microcode malware, processor vendors could introduce an opt-in development mode that allows self-signed updates. Software vendors that want to use such an update in the field, e.g., with processors not in development mode, have to go through a signing process with the \ac{CPU} vendor.

\subsection{Limitations}
\label{cucode:section:discussion:limitations}

At first, we review the limitations of microcode in general. The execution speed of certain computations can be speed up by several orders of magnitude by implementing the algorithm in hardware, e.g., in an ASIC or FPGA. Such performance gains do not apply to computations moved from an x86 implementation to microcode, because essentially it is still software. Merely the decoding is changed, but the resulting operations performed by the functional units of the processor are similar. Furthermore, the intervention of microcode in microarchitectural processes directly implemented in hardware is limited. Custom microcode updates are thus limited to changing the semantics of x86 instructions within the constraints of the existing internal \ac{RISC} instruction set. To the best of our knowledge, no mechanisms exists to periodically trigger an action in microcode to implement an asynchronous monitoring. All actions of custom microcode programs needs to be triggered by an external event. However, as it is possible to intercept arbitrary instructions and microcode-internal processes, there are multiple options to implement a basic form of such a monitoring.

Our microcode research is further limited due to our incomplete knowledge of microcode and the underlying microarchitecture. The information gained through reverse engineering may lack important details or even contain mistakes. This can only be resolved with access to the official documentation of the used features. Our microprograms only run on AMD K8 to K10 family based processors. More modern \acp{CPU} include effective authentication schemes, such as RSA-based public key cryptography, which would need to be bypassed in order to apply a custom update. The microcode update size of the affected \acp{CPU} is limited to 32 triads, which prohibits the implementation of large microprograms. We partly bypassed this restriction by introducing x86 callbacks. However, this bypass is not feasible in scenarios with untrusted operating system kernels such as $\mu$Enclave. More recent \acp{CPU} use larger microcode updates, which is an indication that their patch \ac{RAM} is larger and can potentially accommodate more complex updates. Despite the limited code size on the tested \acp{CPU} no upper bound on the execution time of microcode was encountered and we were able to lock up the \acp{CPU} by forcing it into an endless loop in microcode. Furthermore, we currently can only hook microcoded x86 instructions. Detailed lists of these microcoded instructions for the K8 architecture can be found in~\cite{amd:k8vectorpathlist} at pages 273ff. The instructions listed as VectorPath are microcoded instructions and Direct/DoublePath instructions are decoded in hardware. While there are indications that it is possible to intercept all instructions, our current reverse engineering results do not allow for this. Lastly, the microcode \ac{ROM} readout contains non-correctable read errors induced by dust particles or irregularities. We are currently working on improving the readout and obtaining an error-free version.

\subsection{Correctness of Reverse Engineering Results}

As our results are based on reverse engineering,
we can not guarantee their correctness. Additionally we are limited to observing the output of the \ac{CPU}, any additional details of the microarchitecture such as scheduling or internal state updates are hidden from us.
The observations might constitute unintended behavior of the \ac{CPU} when used outside of its specifications.
However, we verified our conclusions using available resources where possible.
A strong indication that our results are indeed correct
is the fact that we can construct complex microcode programs that behave as expected when executed on the \ac{CPU}.
Additionally the behavior is consistent between \acp{CPU} of the AMD K8 and K10 families,
even though they differ in details such as cache sizes, core counts,
or feature size, and even certain implementation details such as the selection of microcoded instructions.
There are also parallels between our results and the descriptions found in
the patent describing the RISC86 instruction set~\cite{patent:2002:risc86},
which appears to be used internally by the \ac{CPU}.
For example,
the encoding for the conditional codes of microcode jumps are the same as stated in the patent.
We also found similarities in the encoding of individual opcodes,
albeit with differences in length and number of opcode fields.
Lastly, certain operations,
most prominently multiple division variants or steps,
and internal register functions,
e.g.
the address of the next x86 instruction to be executed,
are closely related.
After reconstructing the mapping between virtual and physical microcode
addresses we could also locate the implementation of specific x86 instructions.
By comparing the disassembled microcode with the expected function of the x86 instruction,
we determined that we indeed correctly interpret the bit sequences.
Examples of this are the instructions \texttt{shrd},
whose implementation shows shifts of the argument registers
according to the specifications and the \texttt{wrmsr} opcode,
which at its start has a large number of instructions comparing ECX (the register
number to write to) to specific values consistent with the documented interface.
We also verified individual microcode instructions on their own by copying the bit sequences to a microcode update,
executing them and comparing the output.
This was extended upon during the development of the microcode emulator for which we tested different
input states on both the emulator and the \ac{CPU} to ensure the correctness of our emulation.

A final confirmation of the correctness can be achieved with the cooperation of the \ac{CPU} vendors.
The availability of official specifications and documentation would allow for a faster development
of custom microcode programs and could potentially allow better usage of available \ac{CPU} features.
Unfortunately, we did not receive a response from AMD after we contacted them.

\subsection{Shadow Stacks}
\label{cucode:section:discussion:shadowstacks}

During our research, we considered an opaque shadow stack implementation as a potent use case for a constructive microprogram. However, due to the fact that \texttt{ret} (near, without immediate) is not implemented in microcode, we can not instrument this instruction. As this instruction is a key requirement in implementing an opaque shadow stack, we were unable to create a proof-of-concept. As \ac{CPU} vendors are able to determine the logic on non-microcoded instructions during the design process, they are able to implement such a shadow stack. Below we discuss the advantages of an opaque shadow stack retrofitted by microcode. 

Shadow stack defenses implement a second stack that is kept in sync with the system's default stack. Shadow stacks often possess special properties in order to achieve certain security goals. For example, the shadow stack can be placed in memory that cannot be accessed by normal program instructions~\cite{kuznetsov2014code}, the direction of growth can be inverted to detect illegal stack accesses that yield diverging results~\cite{salamat2008reverse}, or the shadow stack stores only fixed-size elements to preserve control-flow metadata in the event of a stack-based buffer overflow~\cite{clang-safestack}. Shadow stacks ensure the integrity of sensitive data on the stack. Therefore, they are often integrated in code-reuse defenses such as CFI~\cite{dang2015performance,clang-safestack,stackarmor:15,per-input-cfi:2015} in order to protect the backward edge of the control flow. Due to their nature, shadow stack implementations need to extend the logic of instructions operating on the stack such as \texttt{call} and \texttt{ret}. Software-based implementations achieve this by adding instructions at all occurrences during compilation~\cite{kuznetsov2014code,dang2015performance,clang-safestack,gcc-safestack} or with static binary rewriting. In 2015, Davi~\etal~\cite{davi2015hafix} proposed a hardware-assisted shadow stack implementation with low performance overhead. However, the defense still requires the insertion of instructions into the protected application.

Shadow stacks can also be implemented in an opaque way. The semantic of existing stack operations is extended rather than relying on the addition of instructions. Benefits of this approach are compatibility with legacy applications, protection of the whole software stack instead of transformed applications and software libraries only, and potential performance gains due to smaller code size as well as improved utilization of the underlying microarchitecture. Depending on the implementation details, stronger security properties can be enforced, e.g., by placing the shadow stack at a memory area not accessible by conventional user mode instructions. Intel released the specification of \ac{CET} containing a shadow stack in 2016 and added GCC support in 2017~\cite{intel2016cet,gcc2017cet}. However, to date no processor with \ac{CET} support has been released. The \ac{CET} shadow stack is opaque except for some new management instructions such as switch shadow stack. We argue that these management instructions will be microcoded, because they implement complex logic and are not performance critical due to their rare occurrence.

\subsection{Lightweight Syscalls}
\label{cucode:section:syscalls}

The syscall interface is provided by the processor and the operating system to offer services to user space. During its setup, the pointer to the syscall handler in kernel space and the kernel stack pointer are stored in \acp{MSR}. Once the syscall instruction is invoked, the processor reads the corresponding \acp{MSR}, switches the stack, and redirects control flow. The syscall handler then invokes the handler for the requested service according to the given syscall number in register \texttt{eax}. The service handler sanitizes the inputs, checks access privileges (where applicable) and performs its desired action. Ultimately, control is transfered back to user space via the \texttt{sysret} instruction by restoring segment registers, again switching stack and redirecting control to the stored instruction pointer.

The performance overhead imposed by syscalls discourages defenses from invoking them frequently. Thus, vital and critical runtime metadata of defenses are kept in the user space, where they are exposed to attackers. To thwart potential tampering with the metadata, many different kinds of information hiding schemes were introduced in the past years~\cite{kuznetsov2014code,ASLR-Guard,dang2015performance,clang-safestack}. However, information hiding has been shown to be ineffective in several attack scenarios~\cite{gawlik2016enabling,goktacs2016undermining,kollenda2017towards,evans2015missing}. We propose lightweight syscalls implemented in microcode, which are assigned to a dedicated opcode. They leave segment registers, the x86 instruction pointer, and the stack in place. Once the opcode is executed, the microcode implementation switches to kernel mode, performs a desired action, and switches back to user mode. The action is specific to the needs of the particular defense and could for example be a restricted read or write to the defense's metadata in kernel memory. Note that special care must be taken during implementation of the microcode update to not introduce a privilege escalation vulnerability. With lightweight syscalls, defenses such as \ac{CFI} and \ac{CPI} can migrate from information hiding to information isolation enforced by the privilege level of the processor. This can potentially further harden existing defenses against advanced adversaries. Due to the nature of lightweight syscalls, we estimate a low performance overhead. Based on our limited knowledge about microcode, we were unfortunately unable to implement and evaluate such an approach. Future work should explore such a microcode-based defense primitive.

\subsection{Microcode Trojan Detection}
\label{cucode:section:discussion:trojandetection}

Koppe~\etal have shown that microcode updates can contain malicious behavior~\cite{koppe2017reverse}. All presented microcode Trojans rely on the same mechanism to gain initial control, namely the interception of x86 instruction decoding. We found that the interception and the additionally executed micro-ops cause a measurable timing difference. In this paper, we showed that a related technique, namely microcode-assisted instrumentation, already exhibits a measurable performance overhead. Our further tests indicate that even if only a single triad---the smallest possible insertion---is inserted into the logic of an instruction, the overhead can already be measured. Given the unavoidable overhead of switching to the microcode \ac{RAM}, a backdoor inserted via a microcode update is in general detectable.

A detection engine can create a base line by measuring the timing of all instructions with no microcode update applied. Then the engine takes a second measurement with the update under test, compares the results, and reports any timing differences. Note that this method only detects x86 instruction hooks and not necessarily malicious behavior. A malicious update does not always need to insert additional logic into existing instructions, it could, for example, modify the handling of certain, potentially undocumented, \acp{MSR}.

In order to also detect such modifications, the microcode update needs to be decoded and, for example, statically analyzed. Program analysis methods would also consider logic that is not inserted at instruction decoding but other internal processes like exception handling on the microarchitectural level. It is also possible to reason about the Trojan's semantics, thus yielding more accurate results. Trojans (or CPU vulnerabilities that can be exploited as backdoors) can also occur in the microcode \ac{ROM}. The detection of these is more challenging, because their behavior is also contained in the baseline measurement and the \ac{ROM} contents need to be read out to apply static analyses.

However, the same problems that plague traditional malware identification are also applicable to the detection of microcode Trojans. Even if the whole microcode, both \ac{ROM} and \ac{RAM}, is available for analysis, it can be hard to determine if a certain code fragment is benign or malicious in nature. This problem is amplified due the limited understanding of microcode internals. But even access to the full documentation on the subject would not be sufficient, as it is possible to use obfuscation to hide the true nature of a code fragment. Lastly, it would be possible to insert a backdoor outside of the microcode engine and directly change the other functional units of the \ac{CPU}. All-in-all detecting microcode Trojans---or hardware backdoors in general---is a difficult problem in the face of powerful adversaries. %

\subsection{Supporting Newer and Different Architectures}

While we were able to apply our understanding of the K8 architecture
to programming for the K10 architecture, other
architectures are far more difficult to support.
As the K10 is a close evolution of the K8,
the microcode engine remained largely the same.
We mainly noticed differences in the selection of microcoded instructions.
For example, the K10 architecture moved the decoding of all \texttt{ret} instructions to hardware,
while the K8 still performed decoding for some variants of it in microcode.
Moving more instructions to the hardware decoder usually results
in better performance as microcoded decoding takes more time.
During our investigation we also determined that the entry points for microcoded instructions were constant between K8 and K10,
but the implementation then branched to different triads during execution.

The major problem when adapting our findings to new architectures is the strong cryptographic authentication of microcode updates for newer \acp{CPU}.
Only with the ability to execute arbitrary code on the hardware,
it was possible to gain an understanding of the fundamental encoding of microcode~\cite{koppe2017reverse}.
Without such a possibility, any analysis is restricted to interpreting existing code,
usually in the form of microcode updates.
However, even the K8 and K10 architectures use a form of scrambling to obfuscate the plain text of the updates.
Analysis of more modern updates shows that those are most likely protected by strong
cryptographic primitives~\cite{hawkesmicrocode} and thus cannot be analyzed as is.
However, even if the plain text of such an update is acquired,
without a specification or a system to execute the code, it is still challenging to recover the microcode semantics.
Large amounts of data and at least some basic information on the intended functionality of the update would be needed to infer any meaning.
Given the comparatively small size of microcode updates (usually in the range of hundreds of kilobytes for a single \ac{CPU}),
this would probably not be feasible in practice.

Another possibility is the analysis of the microcode \ac{ROM} or engine directly.
Analyzing the engine itself would yield a detailed understanding
of the encoding and available functionality of microcode,
but modern small feature sizes and the high complexity of current \acp{CPU} render this approach difficult.
While reading the \ac{ROM} directly is not as difficult as analyzing a highly optimized microcode engine,
it does not immediately yield the plain text microcode.
As our reverse engineering process showed,
we had to invert multiple permutations of the readout bits in order to obtain the plain text encoding.
This process was heavily dependent on both previous understanding of the
encoding and the ability to execute chosen microcode on the \ac{CPU},
both of which would not be available.
Also there would be no way of verifying the findings,
as the \ac{CPU} would not accept custom updates without the correct signature.
While the public key of the signature could possibly be extracted from the \ac{CPU},
the required private key would only be available to the vendor.
Modifying a single \ac{CPU} via chip editing might resolve this issue,
but such an approach again requires massive hardware reverse engineering efforts
and access to specialized and expensive lab equipment able to operate at the small feature size.
Also such an edit would only allow a single \ac{CPU} to load the custom update,
any unedited \ac{CPU} would refuse it.

In summary, supporting newer \acp{CPU} is mostly prevented by strong authentication of microcode updates.
Once the authentication is circumvented,
e.g.,
by the use of chip editing or side-channel attacks,
our reverse engineering methods can be applied to infer microcode features.
However, vendor support for custom microcode updates is still the
most viable approach to modifying the behavior of \acp{CPU}.

%% file: sections/conclusion.tex
\section{Conclusion}

Vulnerabilities affecting security and safety have accompanied computer systems since their early days. To cope with attacks, numerous defense strategies have been integrated both in software and hardware. In particular, hardware-based defenses implemented with microcode provide increased security and performance, as recently shown by the microcode updates released to address \textsc{Spectre} and \textsc{Meltdown}. However, little is publicly known how security mechanisms are implemented in hitherto closed-source microcode.

In this paper, we demonstrated how modern system security defenses and tools can be implemented in microcode on a modern \ac{COTS} x86 \ac{CPU}. Among others, we provided details how to implement timing attack mitigations, instruction set randomization, and enclave functionality. To this end, we first uncovered new x86 microcode details by a more in-depth hardware reverse engineering and novel strategies to validate the semantics. Finally, we discussed perspectives of customizable microcode and highlighted useful primitives offered by microcode to arm the system security defense landscape.  

In order to foster future research in the area of processor microcode and its applications, we publish the source code of the applications described in this paper as well as the framework used for manipulating and generating microcode~\cite{microcode:amd_microprograms}. We hope this will enable other researchers to extend and build upon our work to design and implement microprograms.

\section*{Acknowledgement}
We thank our shepherd Mathias Payer and the anonymous reviewers for their valuable feedback.
Part of this work was supported by the European Research
Council (ERC) under the European Union's Horizon 2020
research and innovation programme (ERC Starting Grant No.
640110 (BASTION) and ERC Advanced Grant No. 695022 (EPoCH)).
In addition, this work was partly supported by the German Federal Ministry of Education and Research (BMBF Grant 16KIS0592K HWSec and BMBF Grant 16KIS0820 emproof).

\balance

%% file: sections/appendix.tex
\appendix
\onecolumn

\clearpage
\section{Appendix}
\subsection{Hardware Details of the Microcode ROM}
\label{cucode:section:appendix:rom}

\begin{figure*}[!htb]
	\begin{minipage}{\columnwidth}
	\centering
	\resizebox{0.8475\linewidth}{!}{
		\includegraphics{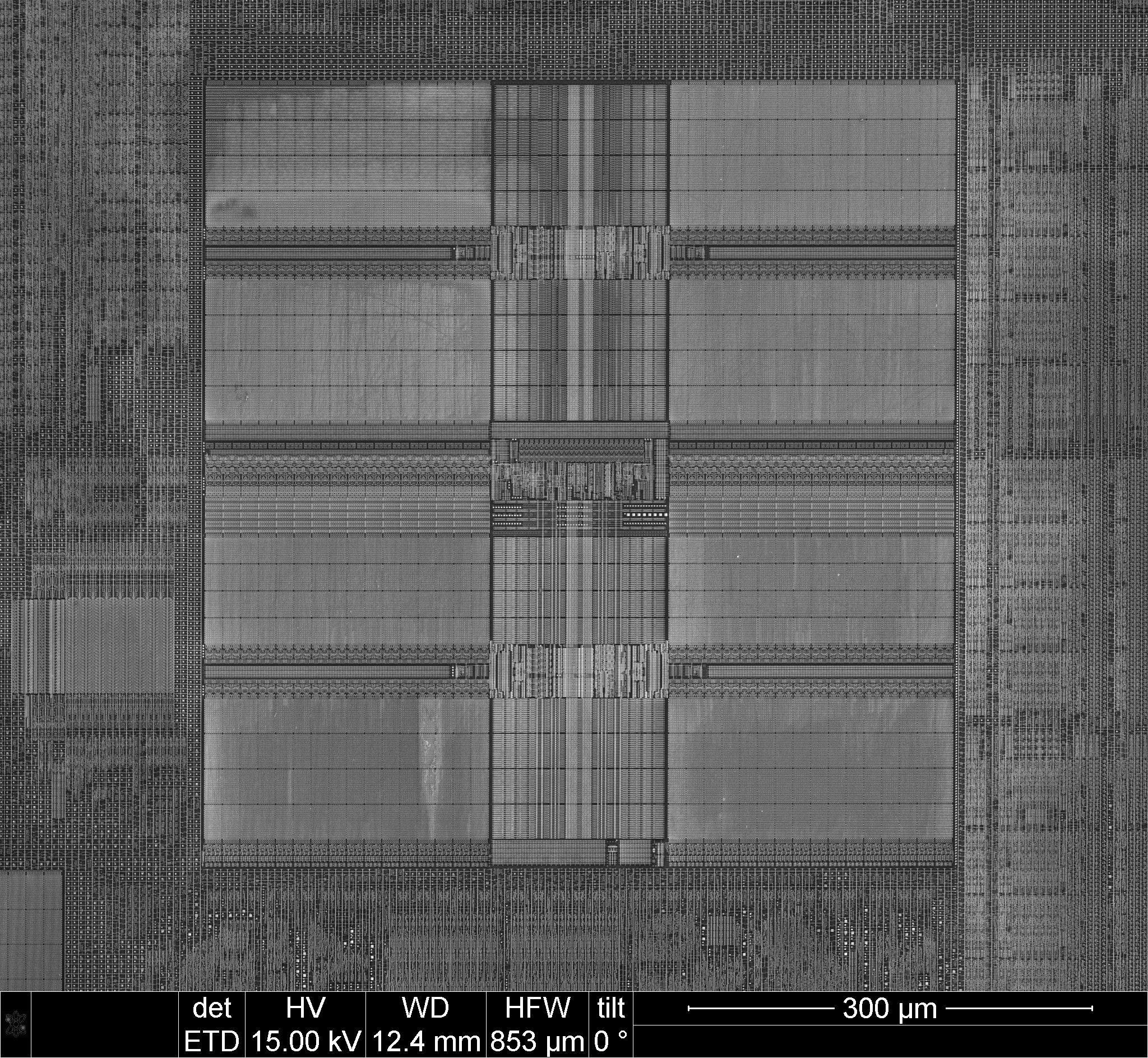}
	}
	\caption{\ac{SEM} image of region R1. The middle part contains the wiring and addressing for the \ac{ROM} and \ac{RAM}. To reduce the average signal path length, the wiring is placed between the two memory areas.}
	\label{cucode:figure:appendix:rom}
	\end{minipage}
\end{figure*}

Figure~\ref{cucode:figure:appendix:rom} shows a \ac{SEM} image of one of the four \acp{ROI}.
As an extension to previous work by Koppe~\etal~\cite{koppe2017reverse},
we further delayered the chip to analyze the region
above the array A2 --- the second array from the bottom.
Its repetitive structure looked visually different compared to the other analyzed NOR-\ac{ROM} arrays.
A cross section and an additional delayering process revealed a prominent structure in the layer underneath, due to which we identified the area as \ac{SRAM}.
Compared to modern \ac{DRAM},
\ac{SRAM} uses more space but can be manufactured in the same process as the adjacent NOR-\ac{ROM}.
Additionally, \ac{SRAM} does not require periodic refreshes to retain the
stored data and is often used in microcontrollers and smaller \acp{SOC}.
The usage of two different storage types in this close proximity
is an indication of a highly optimized in-house design process.
The common practice is to use (third-party) \ac{IP} cores providing a single memory type.

In the \ac{ROM}, the microcode triads are ordered with an eight line interleaving,
meaning that in a linear readout the successor of a triad is found seven triads ahead.
This ordering was verified by searching for all-zero triads at the end of the array A2.
After encountering the first all-zero triad,
more were found at the expected seven triad distance.
Moreover, the hardware layout already hints at the usage of this technique.
Note that these and other techniques used are not implemented for the sake of obfuscating the \ac{ROM} contents,
but instead optimize the storage in regards to die area.

\subsection{\acs{RTL} Representations of Microcode Programs}

In the following, we list the \ac{RTL} form of our custom microcode programs described in the paper.
The \ac{RTL} is the same as used by Koppe~\etal~\cite{koppe2017reverse} and follows the x86 assembly syntax closely.
Where appropriate, the differences to the x86 syntax are highlighted in a comment.
A major difference is the availability of a three operand mode.
In that case, the left-most operand is the destination,
the remaining two operands are the sources.
More examples can be found in our Github repository~\cite{microcode:amd_microprograms}.

{\lstset{language=[x86masm]Assembler, deletekeywords={size}}
\begin{lstlisting}[
	caption={Implementation of our custom \texttt{rdtsc} variant with reduced accuracy.
It completely replaces the default by intercepting triad 0x318,
the entry point for this instruction on the K8 architecture.
The \texttt{dbg} opcode that is used for the read of an internal register sets certain
flags that are not currently supported with standard annotations in the \ac{RTL}.
We omitted the check of the \texttt{CR4.TSD} control bit,
which optionally prevents access to this instruction from usermode.
While we were able to partially reconstruct the check from the \ac{ROM} readout,
we encountered a read error during this and cannot fully and reliably reconstruct the corresponding semantics.
However, this is a limitation of the current state of reverse
engineering and we are working on improving the readout method.},
	label=cucode:listing:rdtsc,captionpos=t
]
; implement default rdtsc semantics, loading TSC to edx:eax
; emit a fixed bitstring, this instruction reads an internal register
dbg 0001010000101111111000000011111111111111110001101010000000001011 
; .q annotation switches to 64 bit operand size
; srl performs a logic shift right
srl.q rdx, t9q, 32
srl.q rax, t9q, 0

; load the and mask
mov t1d, 0xffff
sll t1d, 16
or t1d, 0xff00

; sequence word annotation, continue at the next x86 instruction
; the following triad is still executed after this annotation
.sw_complete

; reduce accuracy of the lower 32 bit TSC
; includes two operations as padding
and eax, t1d
add t2d, 0
add t2d, 0
\end{lstlisting}

\captionof{lstlisting}{Assembly code of a test case for the \ac{ISR}.
The original x86 assembly code is shown on the left. The right side is the translation performed by our transpiler.
Each source instruction maps to a single replacement instruction.
In this case we used a single instruction, \texttt{bound}, to implement all semantics, but it is also possible to repurpose multiple different x86 instructions.
The correct handler is selected by the lower 16 Bits of the displacement given in brackets.
The higher 16 Bits are used as an optional argument for the selected handler.
In the case of memory loads, the argument is the 16 Bit offset of the memory location to be loaded relative to a fixed base address.
The argument to the shift handler is the amount of bits to shift.
The mapping of handler number to semantics is the trivial case in this example: the handler indices are used directly.
However, the full 16 Bits are available to identify handlers.
This also allows for using multiple different indices for the same handler, further strengthening the \ac{ISR}.}
\label{cucode:listing:isr}
\begin{minipage}[t]{.45\textwidth}
    \begin{lstlisting}
mov esi, [msg0]
mov edi, [msg1]
mov ecx, [rc]

add edi, ecx
add esi, edi
mov edi, esi
add esi, esi
shr esi, 8
add esi, edi
    \end{lstlisting}
\end{minipage}\hfill
\begin{minipage}[t]{.45\textwidth}
    \begin{lstlisting}
bound esi, [eax + 0x1]
bound edi, [eax + 0x40001]
bound ecx, [eax + 0x180001]

bound edi, [ecx + 0x4]
bound esi, [edi + 0x4]
bound edi, [esi + 0x0]
bound esi, [esi + 0x4]
bound esi, [eax + 0x80003]
bound esi, [edi + 0x4]
    \end{lstlisting}
\end{minipage}